\documentclass[aps,prb,showpacs,twocolumn,superscriptaddress]{revtex4-1}
\usepackage{bm,color,amsmath,amssymb,mathrsfs,latexsym,graphicx,psfrag,accents,float}
\pdfoutput=1

\newcommand{\ctv}{C_{3v}}

\newcommand{\ctvb}{C_{3v}^{\sma{(b)}}}
\newcommand{\cfv}{C_{4v}}
\newcommand{\csv}{C_{6v}}
\newcommand{\cnv}{C_{nv}}

\newcommand{\cala}{{\cal A}}
\newcommand{\calf}{{\cal F}}

\newcommand{\calc}{{\cal C}}

\newcommand{\gam}[1]{\Gamma_{#1}}

\newcommand{\imp}{\;\;\Rightarrow\;\;}

\newcommand{\bra}[1]{\big\langle#1\big|}
\newcommand{\ket}[1]{\big|#1\big\rangle}

\newcommand{\bea}{\begin{eqnarray}}
\newcommand{\enea}{\end{eqnarray}}
\newcommand{\beq}{\begin{equation}}
\newcommand{\eneq}{\end{equation}}

\newcommand{\pdg}[1]{{#1}^{\phantom{\dagger}}}

\newcommand{\lin}{\notag \\}

\newcommand{\eq}{=&\;}
\newcommand{\ab}{\alpha\beta}

\newcommand{\low}{L$\ddot{\text{o}}$wdin\;}

\newcommand{\W}{{\cal W}}

\newcommand{\bpm}{\begin{pmatrix}}
\newcommand{\epm}{\end{pmatrix}}

\newcommand{\bal}{\begin{align}}
\newcommand{\eal}{\end{align}}

\newcommand{\R}{\mathbb{R}}

\newcommand{\si}{\;\text{sin}\,}

\newcommand{\co}{\;\text{\text{cos}}\,}

\newcommand{\dg}[1]{#1^{\scriptstyle{\dagger}}}

\newcommand{\sma}[1]{\scriptscriptstyle{#1}}

\newcommand{\Z}{\mathbb{Z}}

\newcommand{\mo}{\text{-}1}

\newcommand{\noi}[1]{\noindent (#1)}

\newcommand{\qed}{\nobreak \ifvmode \relax \else
      \ifdim\lastskip<1.5em \hskip-\lastskip
      \hskip1.5em plus0em minus0.5em \fi \nobreak
      \vrule height0.75em width0.5em depth0.25em\fi}

\begin{document}
\title{Spinless Topological Insulators without Time-Reversal Symmetry}
 \author{A. Alexandradinata} \affiliation{Department of Physics, Princeton University, Princeton NJ 08544, USA} 
  \author{Chen Fang}
   \affiliation{Department of Physics, University of Illinois, Urbana IL 61801, USA}
   \affiliation{Micro and Nanotechnology Laboratory, University of Illinois, 208 N. Wright Street, Urbana IL 61801, USA}
  \affiliation{Department of Physics, Princeton University, Princeton NJ 08544, USA}   
\author{Matthew J. Gilbert}
  \affiliation{Department of Electrical and Computer Engineering, University of Illinois, Urbana IL 61801, USA}
  \affiliation{Micro and Nanotechnology Laboratory, University of Illinois, 208 N. Wright Street, Urbana IL 61801, USA}
  \author{B. Andrei Bernevig} \affiliation{Department of Physics, Princeton University, Princeton NJ 08544, USA} 
  

\begin{abstract}
We explore the 32 crystallographic point groups and identify topological phases of matter with robust surface modes. For $n =3,4$ and $6$ of the $\cnv$ groups, we find the first-known 3D topological insulators {without} spin-orbit coupling, and with surface modes that are protected only by point groups, \emph{i.e.}, {not needing} time-reversal symmetry. To describe these $\cnv$ systems, we introduce the notions of (a) a \emph{halved} mirror chirality: an integer invariant which characterizes {half}-mirror-planes in the 3D Brillouin zone, and (b) a \emph{bent} Chern number: the traditional TKNN invariant generalized to bent 2D manifolds. We find that a Weyl semimetallic phase intermediates two gapped phases with distinct halved chiralities. 
\end{abstract}
\date{\today}


\maketitle

\begin{figure}[h]
\centering
\includegraphics[width=8 cm]{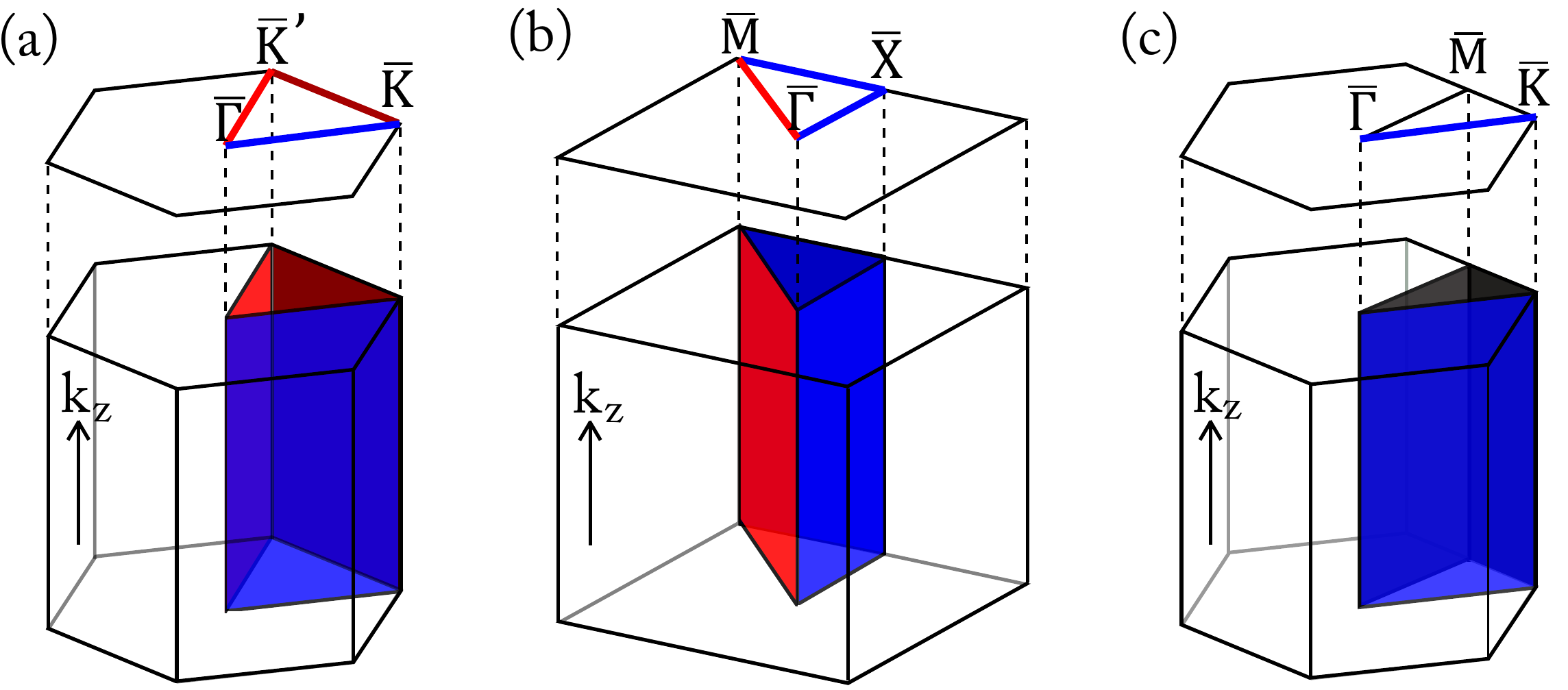}
\caption{ Bottom: (a) Half-mirror-planes (HMP's) in the 3D Brillouin zone (BZ) of a hexagonal lattice with $\ctv^{\sma{(b)}}$ symmetry. Blue face: HMP$_1$. Brown: HMP$_2$. Red: HMP$_3$. (b) HMP's in the 3D BZ of a tetragonal lattice with $\cfv$ symmetry. Red: HMP$_4$. Blue: HMP$_5$. (c) Blue: HMP$_1$ in the 3D BZ of a hexagonal lattice with $\csv$ symmetry. Note that the black-colored submanifold in (c) is not a HMP. In each of (a), (b) and (c), we shall define a bent Chern number on the triangular pipe with its ends identified. Top: Non-black lines are half-mirror-lines (HML's) in the corresponding 2D BZ of the 001 surface; each HML connects two distinct $C_m$-invariant points with $m>2$.}\label{HMP}
\end{figure}

Insulating phases are deemed distinct if they cannot be connected by continuous changes of the Hamiltonian that preserve both the energy gap and the symmetries of the phase; in this sense we say that the symmetry protects the phase. Distinct phases have strikingly different properties -- of experimental interest are the presence of boundary modes, which in many cases distinguish a trivial and a topological phase. The symmetries that are ubiquitous in crystals belong to the space groups, and among them the point groups are the sets of transformations that preserve a spatial point. Despite the large number of space groups in nature, there are few known examples in which boundary modes are protected by crystal symmetries \emph{alone}.\cite{teo2008,Hsieh2012,Xu2012,chaoxingliu2013} In this letter we explore the 32 crystallographic point groups and identify topological phases of matter with robust surface modes. For $n =3,4$ and $6$ of the $\cnv$ groups, we find the first-known 3D topological insulators (TI's) {without} spin-orbit coupling, and with surface modes that are protected only by point groups, \emph{i.e.}, {not needing} time-reversal symmetry (TRS). To describe these $\cnv$ systems, we introduce the notions of (a) a \emph{halved} mirror chirality: an integer invariant which characterizes {half}-mirror-planes in the 3D Brillouin zone (BZ), and (b) a \emph{bent} Chern number: the traditional TKNN invariant\cite{thouless1982} generalized to bent 2D manifolds (illustrated in Fig. \ref{HMP}).

\begin{figure}[h]
\centering
\includegraphics[width=8 cm]{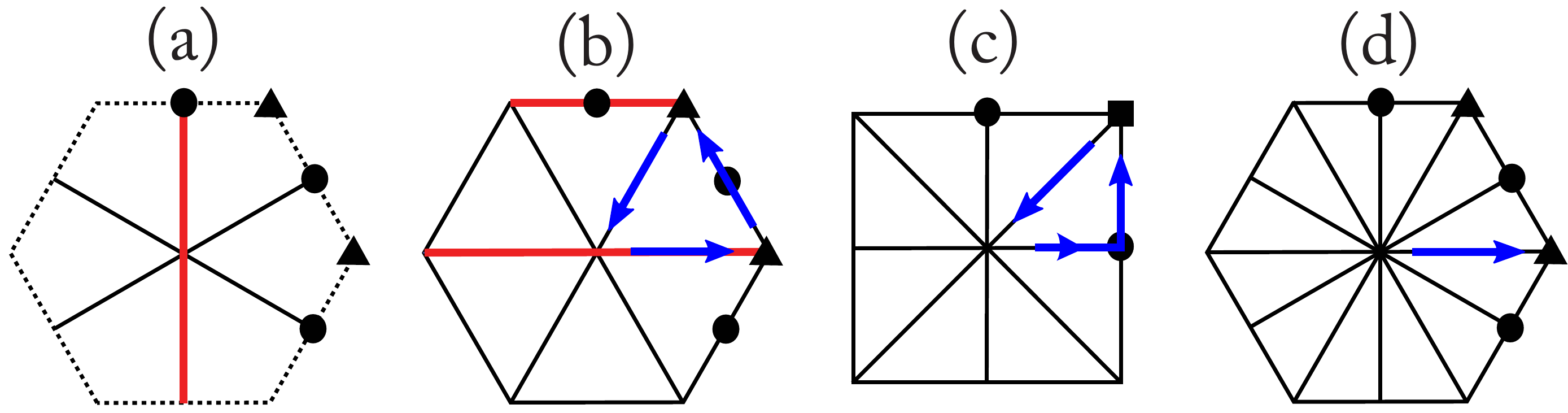}
\caption{Top-down view of 3D BZ's with various symmetries; our line of sight is parallel to the rotational axis. Reflection-invariant planes are indicated by solid lines. The rotationally-symmetric line through $\Gamma$ is present in all cases; all other non-equivalent $C_n$-invariant lines are indicated by circles for $n=2$, triangles for $n=3$ and squares for $n=4$. (a) Hexagonal BZ with $\ctv^{\sma{(a)}}$ symmetry. (b) Hexagonal BZ with $\ctv^{\sma{(b)}}$ symmetry. (c) Tetragonal BZ with $\cfv$ symmetry. (d) Hexagonal BZ with $\csv$ symmetry. For each of $\{\ctv^{\sma{(a)}},\ctv^{\sma{(b)}}\}$, there are two independent mirror Chern numbers, defined as $\calc_e$ ($\calc_o$) in the mirror-even (odd) subspace.\cite{suppprl} In both (a) and (b), $\calc_e$ and $\calc_o$ are defined on a \emph{single} mirror plane indicated in red; in (b) there appear two red lines that correspond to two projected planes; these planes are connected by translation of a reciprocal lattice vector.}\label{mirrorchern}
\end{figure}

To date, all experimentally-realized TI's are strongly spin-orbit-coupled, and a variety of exotic phenomenon originate from this coupling, \emph{e.g.}, Rashba spin-momentum locking on the surface of a TI.\cite{winklerbook} Considerably less attention has been addressed to spinless systems, \emph{i.e.}, insulators and semimetals in which spin-orbit coupling is negligibly weak.\cite{fu2011} The topological classifications of spinless and spin-orbit-coupled systems generically differ. A case in point is SnTe, which is the prototype of a $\cnv$ system with strong spin-orbit coupling.\cite{Hsieh2012} In SnTe, the mirror Chern number\cite{teo2008} was introduced to characterize planes in the 3D BZ which are invariant under reflection, or mirror planes in short. The Bloch wavefunctions in each mirror plane may be decomposed according to their representation under reflection, and each subspace may exhibit a quantum anomalous Hall effect;\cite{Haldane1988} we denote $\calc_e$ ($\calc_o$) as the Chern number in the even (odd) subspace of reflection. One may similarly define mirror Chern numbers for spinless $\cnv$ systems, as illustrated for $\ctv$ in Fig. \ref{mirrorchern}(a) and (b). However, for $n=2,4$ and $6$, such characterization is always trivial due to two-fold rotational symmetry and the lack of spin-orbit coupling, \emph{i.e.}, $\calc_e=\calc_o=0$.\cite{suppprl}

In this work, we propose that point-group-protected surface modes can exist without mirror Chern numbers, if the point group satisfies the following criterion: there must exist at least two high-symmetry points ($\boldsymbol{k_1}$ and $\boldsymbol{k_2}$) in the surface BZ, which admit two-dimensional irreducible representations (irreps) of the little group\cite{tinkhambook} at each point. This is fulfilled by crystals with $\cfv$ and $\csv$ symmetries, but not $\calc_{2v}$. There exist in nature two kinds of $\ctv$: $\ctv^{\sma{(a)}}$ and $\ctv^{\sma{(b)}}$, which differ in the orientation of their mirror planes; compare Fig. \ref{mirrorchern}(a) with (b). Only $\ctv^{\sma{(b)}}$ fulfills our criterion. Henceforth, $\cnv$ is understood to mean $\ctv^{\sma{(b)}}, \cfv$ and $\csv$. We are proposing that surface bands of $\cnv$ systems assume topologically distinct structures on lines which connect $\boldsymbol{k_1}$ to $\boldsymbol{k_2}$. Of all lines that connect $\boldsymbol{k_1}$ to $\boldsymbol{k_2}$, we are particularly interested in half-mirror-lines (HML's), that each satisfies two conditions: (a) It connects two \emph{distinct} $C_m$-invariant points for $m>2$; we illustrate this in Fig. \ref{HMP}, where a $C_m$-invariant point is mapped to itself under an $m$-fold rotation, up to translations by a reciprocal lattice vector. (b) All Bloch wavefunctions in a HML may be diagonalized by a \emph{single} reflection operator. On these HML's, we would like to characterize orbitals that transform in the 2D irrep of $\cnv$, \emph{e.g.}, $(p_x,p_y)$ or $(d_{xz},d_{yz})$ orbitals. We refer to these as the doublet irreps, and all other irreps are of the singlet kind. 

\begin{figure}[h]
\centering
\includegraphics[width=8 cm]{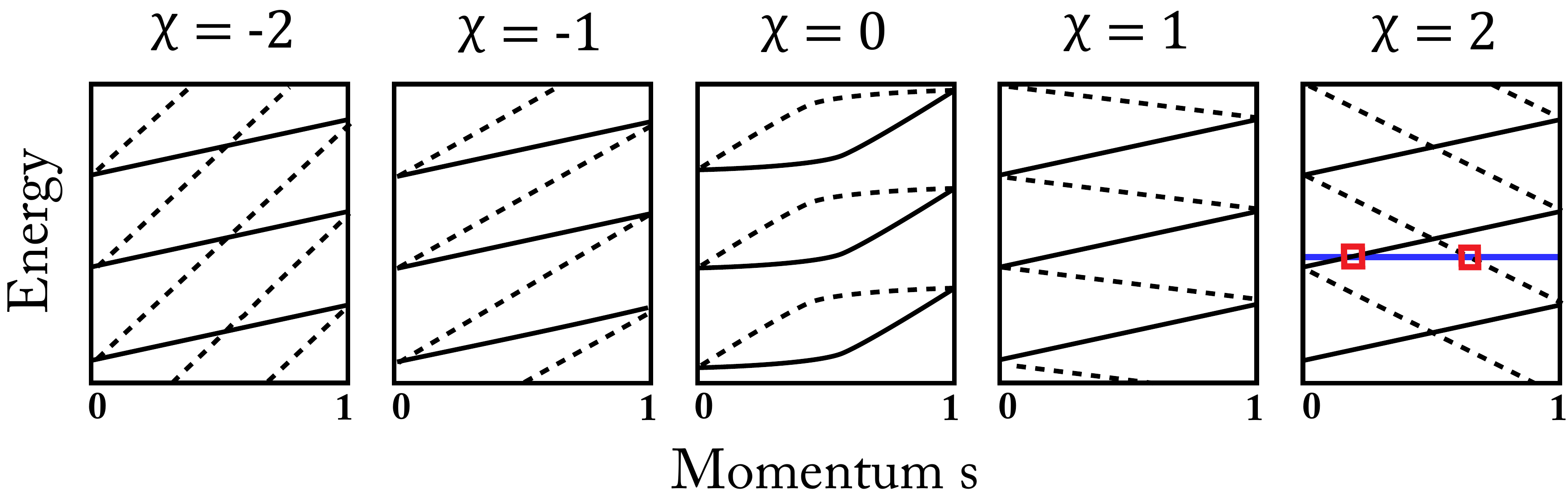}
\caption{ Distinct connectivities of the (001) surface bands along the half-mirror-lines. Black solid (dotted) lines indicate surface bands with eigenvalue $+1$ ($-1$) under reflection $M_i$; crossings between solid and dotted lines are robust due to reflection symmetry. For simplicity, we have depicted all degeneracies at momenta $s=0$ and $s=1$ as dispersing linearly with momentum. This is true if the little group of the wavevector (at $s \in \{0,1\}$) is $\ctv$, but for $\cfv$ and $\csv$ such crossings are in reality quadratic.\cite{suppprl}}\label{connectivity}
\end{figure}

We begin by parametrizing HML$_i$ with $s_i \in [0,1]$, where $s_i=0$ $(1)$ at the first (second) $C_m$-invariant point. The subscript $i$ labels the different HML's in a $C_{nv}$ system; the $i$'th HML is invariant under a specific reflection $M_i$. At $s_i=0$ and $1$, (001) surface bands form doubly-degenerate pairs with opposite mirror eigenvalues, irrespective of whether the system has TRS. To prove this, let $U(g)$ represent the symmetry element $g$ in the orbital basis. Suppose $U(M_i)\ket{\eta} = \eta\ket{\eta}$ for $\eta \in \{\pm 1\}$. By assumption, $\ket{\eta}$ transforms in the doublet representation, \emph{i.e.}, it is a linear combination of states with complex eigenvalues under $U(C_m)$, for $m>2$. It follows that $\big[\,U(C_m)-U(C_m^{\mo})\,\big]\ket{\eta}$ is not a null vector, and moreover it must have mirror eigenvalue $-\eta$ due to the relation $M_i\,C_m\,M_i^{\mo} = C_m^{\mo}$. Given these constraints at $s_i=0$ and $1$, there are $\Z$ ways to connect mirror-even bands to mirror-odd bands, as illustrated schematically in Fig. \ref{connectivity}. We define the halved mirror chirality $\pdg{\chi}_i \in \Z$ as the \emph{difference} in number of mirror-even chiral modes  with mirror-odd chiral modes; if $\pdg{\chi}_i \neq 0$, the surface bands robustly interpolate across the energy gap. $\pdg{\chi}_i$ may be easily extracted by inspection of the surface energy-momentum dispersion: first draw a constant-energy line within the bulk energy gap and parallel to the HML, \emph{e.g.}, the blue line in Fig. \ref{connectivity}. At each intersection with a surface band, we calculate the sign of the group velocity $dE/ds_i$, and multiply it with the eigenvalue under reflection $M_i$. Finally, we sum this quantity over all intersections along HML$_i$ to obtain $\pdg{\chi}_i$. In Fig. \ref{connectivity}, we find two intersections as indicated by red squares, and $\pdg{\chi}_i = (1)(1) + (-1)(-1)=2$. The $\Z$-classification of (001) surface bands relies on doublet irreps in the surface BZ; on surfaces which break $\cnv$ symmetry, the surface bands transform in the singlet irreps, and cannot assume topologically distinct structures. 

Thus far we have described the halved chirality $\pdg{\chi}_i$ as a topological property of surface bands along HML$_i$, but we have not addressed how $\pdg{\chi}_i$ is encoded in the \emph{bulk} wavefunctions. Taking $\hat{z}$ to lie along the rotational axis, each HML$_i$ in the surface BZ is the $\hat{z}$-projection of a half-mirror-plane (HMP$_i$) in the 3D BZ, as illustrated in Fig. \ref{HMP}. Each HMP connects two \emph{distinct} $C_m$-invariant lines for $m>2$, and all Bloch wavefunctions in a HMP may be diagonalized by a single reflection operator. HMP$_i$ is parametrized by $t_i \in [0,1]$ and $k_z \in (-\pi,\pi]$, where $t_i=0$ $(1)$ along the first (second) $C_m$-invariant line. Then the halved mirror chirality has the following expression by bulk wavefunctions:\cite{suppprl}
\bal \label{mirrorchirality}
\pdg{\chi}_i = \frac{1}{2\pi}\,\int_{\text{HMP}_i} dt_i\,dk_z\,(\,\calf_{e} - \calf_{o}\,) \in \Z.
\end{align}
For spinless representations, $M_i^2 = I$, and we label bands with mirror eigenvalue $+1$ $(-1)$ as mirror-even (mirror-odd). $\calf_{e}$ ($\calf_{o}$) is defined as the Berry curvature of occupied doublet bands,\cite{berry1984,zak1982,*zak1989} as contributed by the mirror-even (-odd) subspace. 

For $\ctv^{\sma{(b)}}$, there exists three independent HMP's as illustrated in Fig. \ref{HMP}-(a), which we label by $i \in \{1,2,3\}$; all other HMP's are related to these three by symmetry. The $\csv$ group is obtained from $\ctv^{\sma{(b)}}$ by adding six-fold rotational symmetry, which enforces $\pdg{\chi}_2=0$, and $\pdg{\chi}_1=- \pdg{\chi}_3$. The sign in the last identity is fixed by our parametrization of $\{t_i\}$, which increase in the directions indicated by blue arrows in Fig. \ref{mirrorchern}. Thus, HMP$_1$ is the sole independent HMP for $C_{6v}$.  Finally, we find that there are two HMP's for $C_{4v}$, labelled by $i \in \{4,5\}$ (Fig. \ref{HMP}(b)). Unlike the other highlighted HMP's, HMP$_5$ is the union of two mirror faces, HMP$_{5a}$ and HMP$_{5b}$, which are related by a $\pi/2$ rotation. States in HMP$_{5a}$ are invariant under the reflection $M_y: y \rightarrow -y$, while in HMP$_{5b}$ the relevant reflection is $M_x: x \rightarrow -x$. The product of these orthogonal reflections is a $\pi$ rotation ($C_2$) about $\hat{z}$, thus $M_x = C_2\,M_y$. In the doublet representation, all orbitals are odd under a $\pi$-rotation, thus $U(C_2) = -I$ and all states in HMP$_5$ may be labelled by a \emph{single} operator $M_y \equiv M_5$. 

\begin{figure}[h]
\centering
\includegraphics[width=8 cm]{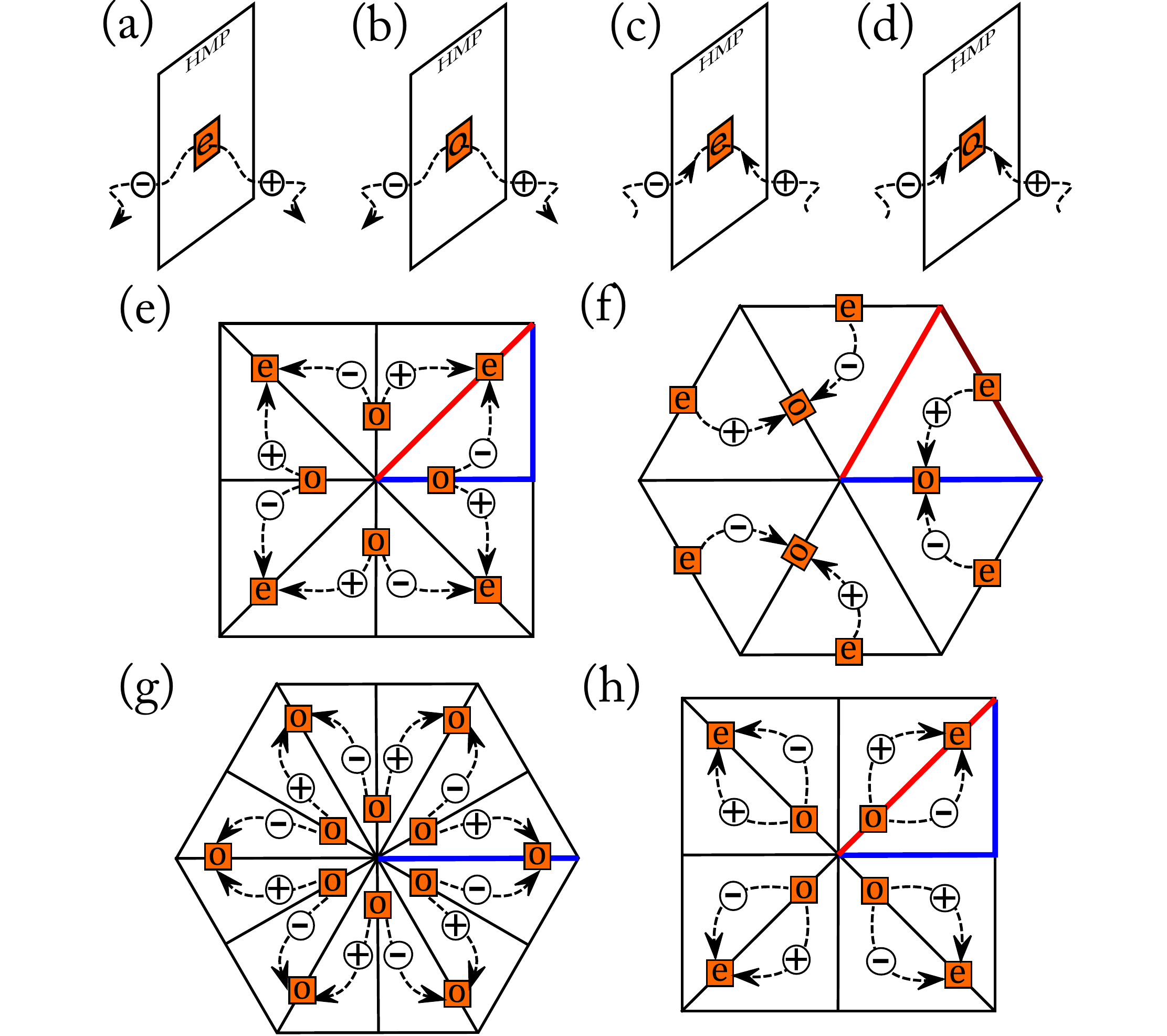}
\caption{(a) to (d) illustrate how the halved chirality of a HMP may change. The direction of arrows indicate whether Weyl nodes are created or annihilated. + (-) labels a Berry monopole with positive (negative) charge; $e$ ($o$) labels a crossing in the HMP between mirror-even (-odd) bands. (e) to (h): In four examples, we provide a top-down view of the trajectories of Weyl nodes, in the transition between two distinct gapped phases. Our line of sight is parallel to $\hat{z}$.  Black lines indicate mirror faces, while colored lines specially indicate HMP's, with the same color legend as in Fig. \ref{HMP}. (e) and (h) describe $\cfv$ systems, (f) $\ctv^{\sma{(b)}}$ and (g) $\csv$.}\label{weyltrajectories}
\end{figure}

The invariants $\{\pdg{\chi}_i\}$ are well-defined so long as bulk states in the HMP's are gapped, which is true of $C_{nv}$ insulators. These invariants may also be used to characterize $\cnv$ semimetals, so long as the gaps close away from the HMP's. Such band touchings are generically Weyl nodes,\cite{Murakami2007B,wan2010} though there exist high-symmetry exceptions.\cite{burkov2011} The chirality of each Weyl node is its Berry charge, which is positive (negative) if the node is a source (sink) of Berry flux. By the Nielsen-Ninomiya theorem, the net chirality of all Weyl nodes in the BZ is zero.\cite{NIELSEN1981} To make progress, we divide the BZ into `unit cells', such that the properties of one `unit cell' determine all others by symmetry. As seen in Fig. \ref{HMP}, these `unit cells' resemble the interior of triangular pipes; they are known as orbifolds, or the quotient spaces $T^3/\cnv$. The net chirality of an orbifold \emph{can} be nonzero, and is determined by the Chern number on the 2D boundary of the orbifold. As each boundary resembles the surface of a triangular pipe, we call it a bent Chern number. We define $\calc_{123}, \calc_{45}$ and $\calc_6$ as bent Chern numbers for $\ctv^{\sma{(b)}}, \cfv$ and $\csv$ respectively. $\ctv^{\sma{(b)}}$ systems in the doublet representation are described by four invariants $(\pdg{\chi}_{1}, \pdg{\chi}_{2}, \pdg{\chi}_{3}, \calc_{123})$, which are related by  $\text{parity}[ \,\pdg{\chi}_{1} + \pdg{\chi}_{2} + \pdg{\chi}_{3}\,] = \text{parity}[ \,\calc_{123}\,].$\cite{suppprl} If $\sum_{i=1}^3 \pdg{\chi}_{i}$ is odd, $\calc_{123}$ must be nonzero due to an odd number of Weyl nodes within the orbifold, which implies the system is gapless. If the system is gapped, $\sum_{i=1}^3 \pdg{\chi}_{i}$ must be even. However, the converse is not implied.  For $\cfv$, a similar relation holds: $\text{parity}[ \,\pdg{\chi}_{4} + \pdg{\chi}_{5}\,] = \text{parity}[\, \calc_{45}\,].$\cite{suppprl} These parity constraints may be understood in light of a Weyl semimetallic phase that intermediates two gapped phases with distinct halved chiralities. There are four types of events that alter the halved chirality $\pdg{\chi}_i$ of HMP$_i$; we explain how Weyl nodes naturally emerge in the process. (i) Suppose the gap closes between two mirror-even bands in HMP$_i$. Around this band touching, bands disperse linearly within the mirror plane, and quadratically in the direction orthogonal to the plane. Within HMP$_i$, the linearized Hamiltonian around the band crossing describes a massless Dirac fermion in the even representation of reflection. If the mass of the fermion inverts sign, $\int_{HMP_i} \calf_{e}/2\pi$ changes by $\eta \in \{\pm 1\}$, implying that $\pdg{\chi}_i$ also changes by $\eta$ through (\ref{mirrorchirality}). This quantized addition of Berry flux is explained by a splitting of the band-touching into two Weyl nodes of opposite chirality, and on opposite sides of the mirror face (Fig. \ref{weyltrajectories}(a)). In analogy with magnetostatics, the initial band touching describes the nucleation of a dipole, which eventually splits into two opposite-charge monopoles; the flux through a plane separating two monopoles is unity. (ii) The same argument applies to the splitting of dipoles in the mirror-odd subspace, which alters $\int_{HMP_i} \calf_{o}/2\pi$ by $\kappa \in \{\pm 1\}$, and  $\pdg{\chi}_i$ by $-\kappa$. For (iii) and (iv), consider two opposite-charge monopoles which converge on HMP$_i$ and annihilate, causing $\pdg{\chi}_i$ to change by unity. The sign of this change is determined by whether the annihilation occurs in the mirror-even or odd subspace. The transition between two distinct gapped phases is characterized by a transfer of Berry charge, either between two distinct HMP's (Fig. \ref{weyltrajectories}(e)-(g)), or within the same HMP and occuring between two different representations of reflection (Fig. \ref{weyltrajectories}(h)). In the intermediate semimetallic phase, the experimental implications include Fermi arcs on the (001) surface.\cite{wan2010,halasz2012,chen2012}

\begin{figure}[h]
\centering
\includegraphics[width=8.5 cm]{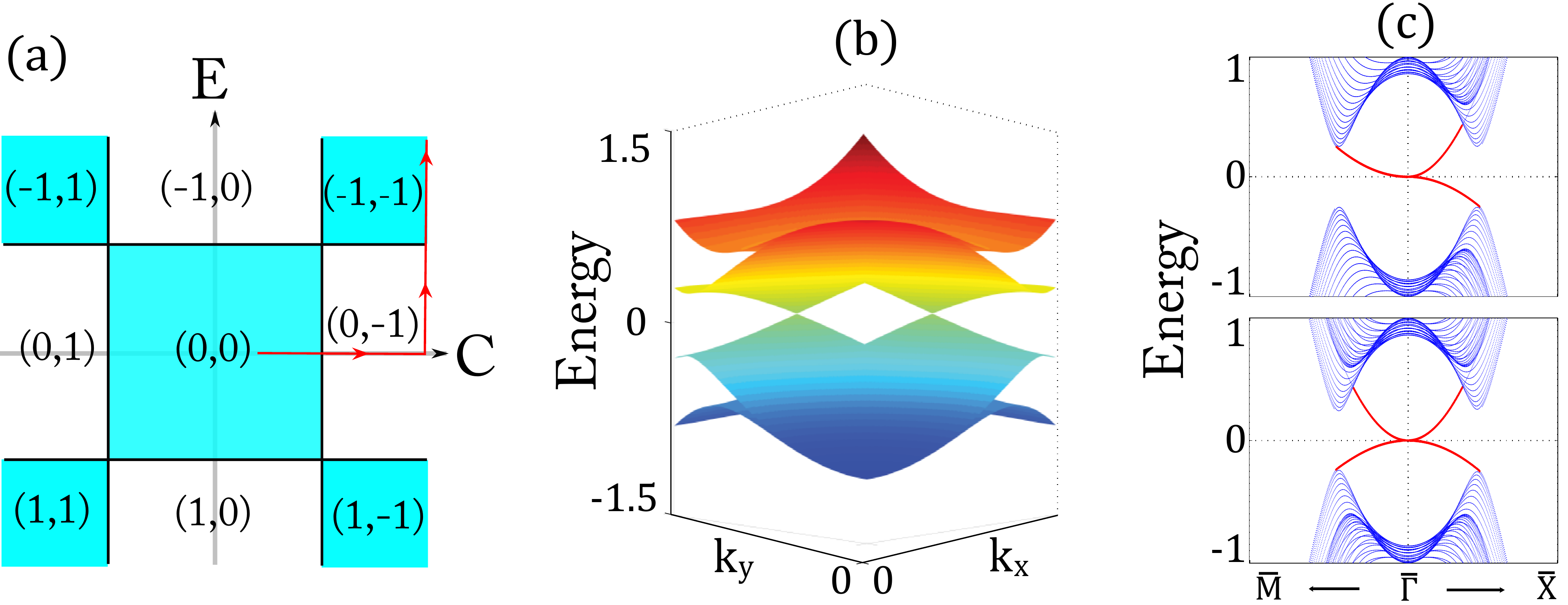}
\caption{(a) Phase diagram of $\cfv$ model (\ref{cfvmodel}); $C$ and $E$ are varied  to induce phase transitions. Blue (uncolored) regions correspond to gapped (gapless) phases. The halved chiralities in each phase are indicated by two integers: $({\chi}_4,{\chi}_5)$. The blue square in the center is approximately bounded by $|C|<2$ and $|E|<2$. (b) Bulk dispersion of the semimetallic phase; $({\chi}_4,{\chi}_5)=(0,-1)$. (c) Surface dispersions along $\bar{M}-\bar{\Gamma}-\bar{X}$. Top of (c): semimetal with $({\chi}_4,{\chi}_5)=(0,-1)$. Bottom: TI with $({\chi}_4,{\chi}_5)=(-1,-1)$.}\label{fig:c4vmodel}
\end{figure}

To exemplify our theory, we consider a $\cfv$ model on a tetragonal lattice, which comprises two interpenetrating cubic sublattices; additional models with $\ctv^{\sma{(b)}}$ and $\csv$ symmetries are detailed in the Supplementary Material. The Bloch Hamiltonian is
\bal \label{cfvmodel}
& H(\boldsymbol{k}) =  \big[\,-1 + 8\,f_1(\boldsymbol{k})\,\big]\,\Gamma_{03} + C\,f_2(\boldsymbol{k})\,\gam{01} \lin
&\;+2\,f_3(\boldsymbol{k})\,\gam{11}  +  E\,f_4(\boldsymbol{k})\,\gam{32} + 2\,f_5(\boldsymbol{k})\,\gam{12},
\end{align}
where $f_1 = 3-\text{cos}(k_x)-\text{cos}(k_y)-\text{cos}(k_z)$, $f_2 = \text{cos}(k_y)-\text{cos}(k_x)$, $f_3 = 2-\text{cos}(k_x)-\text{cos}(k_y)$, $f_4 = \text{sin}(k_x)\,\text{sin}(k_y)$ and $f_5 = \text{sin}(k_z)$.
We define $\Gamma_{ab} = \sigma_a \otimes\tau_b$, where $\sigma_i$ and $\tau_i$ are Pauli matrices for $i \in \{1,2,3\}$; $\sigma_0$ and $\tau_0$ are identities in each 2D subspace.  $\ket{\sigma_3= \pm 1,\tau_3=+1}$  label $\{p_x \pm i p_y\}$ orbitals on one sublattice, and $\ket{\sigma_3= \pm 1,\tau_3=-1}$  label $\{p_x \mp i p_y\}$ orbitals on the other. This Hamiltonian has the four-fold symmetry: $\Gamma_{33}\,H(k_x,k_y,k_z)\,\Gamma_{33} = H(\,-k_y,k_x,k_z\,)$, and the reflection symmetry: $\Gamma_{10} \,H(k_x,k_y,k_z)\,\gam{10} = H(k_x,-k_y,k_z)$. A phase diagram is plotted in Fig. \ref{fig:c4vmodel}(a) for different parametrizations of (\ref{cfvmodel}); the sweep of parameters indicated by the red line produces the Weyl trajectories of Fig. \ref{weyltrajectories}(e). In Fig. \ref{fig:c4vmodel}(b) and (c) we illustrate the energy dispersions at two points along this sweep. 

We hope our classification stimulates a search for materials with $\cnv$ symmetry. Why is $\cnv$ special? We propose sufficient criteria for gapless surface modes whose robustness rely on a symmetry group. Minimally, (i) the symmetry must be unbroken by the presence of the surface. Additionally, either (ii) a reflection symmetry exists so that mirror-subspaces can display a quantum anomalous Hall effect, or (iii) there exist at least two high-symmetry points in the surface BZ, which admit higher-than-one dimensional irreps of the symmetry group. In addition to predicting new topological materials, these criteria are also satisfied by the well-known SnTe class,\cite{Hsieh2012} and also the $\Z_2$ insulators.\cite{fu2006,kane2005A,kane2005B,bernevig2006a,bernevig2006c,koenig2007,fu2007b,moore2007,roy2009,*roy2009a,hsieh2008} Among the $32$ crystallographic point groups, only the $C_n$ and $\cnv$ groups are preserved for a surface that is orthogonal to the rotational axis.\cite{tinkhambook} Though all $\cnv$ groups satisfy (ii), the lack of spin-orbit coupling implies only $\ctv^{\sma{(a)}}$ and $\ctv^{\sma{(b)}}$ systems can have nonvanishing mirror Chern numbers.\cite{suppprl} Finally, only a subset of the $\cnv$ groups possess two-dimensional irreps which satisfy (iii): $\ctv^{\sma{(b)}}, \cfv$ and $\csv$. In the second row of Tab. \ref{cnvsum}, we list the topological invariants which characterize $\ctv,\cfv$ and $\csv$, for bands of any irrep. In addition to the well-known mirror Chern numbers $(\calc_e,\calc_o)$, we have introduced the bent Chern numbers as a measure of the Berry charge in each orbifold $T^3/\cnv$. If the orbital character of bands near the gap is dominated by the doublet irreps, then the halved mirror chirality  $\pdg{\chi}_i$ becomes a useful characterization, as seen in the third row of Tab. \ref{cnvsum}. In particular, the mirror Chern numbers of $\ctv^{\sma{(b)}}$ are completely determined by $(\pdg{\chi}_{1}, \pdg{\chi}_{2}, \pdg{\chi}_{3}, \calc_{123})$.\cite{suppprl} The singlet and doublet irreps of realistic systems are often hybridized. The topological surface bands that we predict here are robust, so long as this hybridization does not close the bulk gap, and if there are no errant \emph{singlet} surface bands within the gap.\cite{fu2011} 

\vspace*{-5pt}
\begin{table}[h]
		\caption{Topological classification of spinless insulators and semimetals with $\cnv$ symmetry. \label{cnvsum}}	
	\centering
		\begin{tabular} {c c c c} \hline
			  $\ctv^{\sma{(a)}}$ & $\ctv^{\sma{(b)}}$ & $\cfv$ & $\csv$  \\ \hline 
			  	$\;\;\;\;\calc_e,\calc_o\;\;\;\;$ & $\calc_e,\calc_o$ & $\calc_{45}$ & $\;\;\;\;\calc_{6}\;\;\;\;$ 	\\
			$\;\;\;\;\calc_e,\calc_o\;\;\;\;$ & $\;\;\pdg{\chi}_{1}, \pdg{\chi}_{2}, \pdg{\chi}_{3}, \calc_{123}\;\;$ & $\;\;\pdg{\chi}_{4}, \pdg{\chi}_{5}, \calc_{45}\;\;$ & $\;\;\;\;\pdg{\chi}_{1}, \calc_{6}\;\;\;\;$	
		\end{tabular}	
\end{table}

To conclude, we discuss generalizations of our findings. In addition to materials whose full group is $\cnv$, we are also interested in higher-symmetry materials whose point groups reduce to $\cnv$ subgroups in the presence of a surface, for $n=3,4$ or $6$. We insist that these higher-symmetry point groups have neither (a) a reflection plane that is orthogonal to the principal $C_n$ axis, nor (b) a two-fold axis that lies perpendicular to the $C_n$ axis, and parallel to the mirror-plane. The presence of either (a) or (b) imposes $\chi=\calc_e=\calc_o=0$ in any (half) mirror-plane of the $\cnv$ system.\cite{suppprl} Only one such higher-symmetry point group exists: $D_{3d}$ reduces to $\ctv$ on the 111 surface.\cite{kosterbook} Many $\cnv$ systems naturally have TRS, which  constrains all $\ctv^{\sma{(a)}}$ and $\ctv^{\sma{(b)}}$ invariants in Tab. \ref{cnvsum} to vanish, with one independent exception for $\ctv^{\sma{(b)}}$: $\pdg{\chi}_1=-\pdg{\chi}_3$ can be nonzero.\cite{suppprl} TRS does not constrain the invariants of $\cfv$ or $\csv$. Our analysis of charge-conserving systems are readily generalized to spinless superconductors which are describable by mean-field theory. Due to the particle-hole redundancy of the mean-field Hamiltonian, the only nonvanishing invariants from Tab. \ref{cnvsum} belong to $\ctv^{\sma{(a)}}$ and $\ctv^{\sma{(b)}}$; among these nonvanishing invariants, the only constraint is $\pdg{\chi}_1=\pdg{\chi}_3$ for $\ctv^{\sma{(b)}}$.\cite{suppprl} While we have confined our description to electronic systems without spin-orbit coupling, the halved chirality is generalizable to photonic crystals which are inherently spinless, and also to cold atoms. Finally, we point out that the bent Chern number and the halved chirality are also valid characterizations of spin-orbit-coupled systems with mirror symmetry. In particular, (\ref{mirrorchirality}) applies to representations with spin if we redefine the mirror-even (-odd) bands as having mirror eigenvalues $+i$ ($-i$).\cite{suppprl} The implications are left to future work.

\emph{Acknowledgements}
AA is grateful to Yang-Le Wu and Tim Lou for insightful interpretations of this work. AA and BAB were supported by Packard Foundation, 339-4063--Keck Foundation, NSF CAREER DMR-095242, ONR - N00014\text{-}11\text{-}1-0635 and  NSF-MRSEC DMR-0819860 at Princeton University. This work was also supported by DARPA under SPAWAR Grant No.: N66001\text{-}11\text{-}1-4110. CF is supported by ONR-N0014-11-1-0728 for salary and DARPA-N66001-11-1-4110 for travel. MJG acknowledges support from the ONR under grant N0014-11-1-0728 and N0014-14-1-0123, a fellowship from the Center for Advanced Study (CAS) at the University of Illinois.


\appendix

\begin{widetext}

The Supplementary Material is organized in the following manner. In App. \ref{app:symminTB}, we describe how point-group symmetries constrain a tight-binding Hamiltonian, and introduce the notion of little groups. In App. \ref{app:vanishmc}, we formulate the mirror Chern numbers in systems with $\cnv$ symmetry. For electronic systems with negligible spin-orbit coupling, or intrinsically spinless systems, we  show that the mirror Chern numbers must vanish for $C_{2v},\cfv$ and $\csv$. In App. \ref{app:kdotp}, we identify the two-dimensional irreducible representations (irreps) of the $\cnv$ groups, then derive minimal-derivative, effective Hamiltonians for these doublet irreps. In App. \ref{app:halved}, we formulate the halved mirror chirality ($\chi$), and prove that it is quantized to integers. Then we identify certain symmetries beyond the $\cnv$ group which constrain $\chi$ to vanish. In App. \ref{app:relatechibent}, we formulate the bent Chern numbers, and show how they are related to the mirror Chern numbers and the halved chiralities. In App. \ref{app:models}, we describe models with nontrivial $\chi$ for the $\ctvb$ and $\csv$ groups. Finally, in App. \ref{app:trs} (App. \ref{app:sc}), we explain the role of time-reversal symmetry (particle-hole redundancy) in constraining some of these topological invariants. The last section is applicable to spinless superconductors with $\cnv$ symmetry.

\section{Point-group symmetry in tight-binding Hamiltonians} \label{app:symminTB}

\subsection{Review of the tight-binding Hamiltonian}

In the tight-binding method, the Hilbert space is reduced to a finite number of \low orbitals $\varphi_{\boldsymbol{R},\alpha}$, for each unit cell labelled by the Bravais lattice (BL) vector $\boldsymbol{R}$.\cite{slater1954,goringe1997,lowdin1950} In Hamiltonians with discrete translational symmetry, our basis vectors are
\bal \label{basisvec}
\phi_{\boldsymbol{k}, \alpha}(\boldsymbol{r}) = \tfrac{1}{\sqrt{N}} \sum_{\boldsymbol{R}} e^{i\boldsymbol{k} \cdot (\boldsymbol{R}+\boldsymbol{r_{\alpha}})} \pdg{\varphi}_{\boldsymbol{R},\alpha}(\boldsymbol{r}-\boldsymbol{R}-\boldsymbol{r_{\alpha}}),
\end{align}
which are periodic in lattice translations $\boldsymbol{R}$. $\boldsymbol{k}$ is a crystal momentum, $N$ is the number of unit cells, $\alpha$ labels the \low orbital, and $\boldsymbol{r_{\alpha}}$ denotes the position of the orbital $\alpha$ as measured from the origin in each unit cell. The tight-binding Hamiltonian is defined as 
\bal
H(\boldsymbol{k})_{\alpha \beta} = \int d^dr\,\phi_{\boldsymbol{k},\alpha}(\boldsymbol{r})^* \,\hat{H} \,\phi_{\boldsymbol{k},\beta}(\boldsymbol{r}), 
\end{align}
where $\hat{H} = p^2/2m + V(\boldsymbol{r})$ is the single-particle Hamiltonian. The energy eigenstates are labelled by a band index $n$, and defined as $\psi_{n,\boldsymbol{k}}(\boldsymbol{r}) = \sum_{\alpha}  \,u_{n,\boldsymbol{k}}(\alpha)\,\phi_{\boldsymbol{k}, \alpha}(\boldsymbol{r})$, where  
\bal
\sum_{\beta} H(\boldsymbol{k})_{\ab} \,u_{n,\boldsymbol{k}}(\beta)  = \varepsilon_{n,\boldsymbol{k}}\,u_{n,\boldsymbol{k}}(\alpha).
\end{align}
We employ the braket notation:
\bal
H(\boldsymbol{k})\,\ket{u_{n,\boldsymbol{k}}} = \varepsilon_{n,\boldsymbol{k}}\,\ket{u_{n,\boldsymbol{k}}}.
\end{align}
Due to the spatial embedding of the orbitals, the basis vectors $\phi_{\boldsymbol{k},\alpha}$ are generally not periodic under $\boldsymbol{k} \rightarrow \boldsymbol{k}+\boldsymbol{G}$ for a reciprocal lattice (RL) vector $\boldsymbol{G}$. This implies that the tight-binding Hamiltonian satisfies:
\bal \label{aperiodic}
H(\boldsymbol{k}+\boldsymbol{G}) = V(\boldsymbol{G})^{\mo}\,H(\boldsymbol{k})\,V(\boldsymbol{G}),
\end{align}
where $V(\boldsymbol{G})$ is a unitary matrix with elements: $[V(\boldsymbol{G})]_{\ab} = \delta_{\ab}\,e^{i\boldsymbol{G}\cdot \boldsymbol{r_{\alpha}}}$. 

\subsection{Symmetry constraints of the tight-binding Hamiltonian} \label{app:rep}

We define the creation operator for a \low function $\varphi_{\boldsymbol{R},\alpha}$ as $\dg{c}_{\alpha}(\boldsymbol{R}+\boldsymbol{r_{\alpha}})$. From (\ref{basisvec}), the creation operator for a Bloch basis vector $\phi_{\boldsymbol{k},\alpha}$ is
\bal \label{blochbasis}
\dg{c}_{\boldsymbol{k}, \alpha} = \tfrac{1}{\sqrt{N}} \sum_{\boldsymbol{R}} \,e^{i\boldsymbol{k} \cdot (\boldsymbol{R} + \boldsymbol{r_{\alpha}})}\,\dg{c}_{\alpha}(\boldsymbol{R}+\boldsymbol{r_{\alpha}}).
\end{align}
Let $g$ be a point-group element that is represented by $D(g)$ in $\R^d$, and by $U(g)$ in the basis of \low orbitals:
\bal
\hat{g}\,\dg{c}_{\alpha}(\boldsymbol{R}+\boldsymbol{r_{\alpha}})\,\hat{g}^{\mo} = \dg{c}_{\beta}\big(\,D(g)\boldsymbol{R}+\boldsymbol{\Delta}_{\beta \alpha}+\boldsymbol{r_{\beta}}\,\big) \,U_{\beta \alpha}.
\end{align}
$D(g)$ is orthogonal: $D(g)^t= D(g)^{\mo}$, and we have defined $\boldsymbol{\Delta}_{\beta \alpha} = D(g)\boldsymbol{r_{\alpha}}-\boldsymbol{r_{\beta}}$. A Bravais lattice (BL) that is symmetric under $g$ satisfies two conditions:

\noi{i}  for any BL vector $\boldsymbol{R}$, $D(g)\boldsymbol{R}$ is also a BL vector: 
\bal \label{pgscond1}
\forall \boldsymbol{R} \in \text{BL}, \;\; D(g)\boldsymbol{R} \in \text{BL}.
\end{align}
\noi{ii} If $g$ transforms an orbital of type $\alpha$ to another of type $\beta$, \emph{i.e.}, $U(g)_{\beta \alpha}$ is nonzero, then $D(g)(\boldsymbol{R}+\boldsymbol{r_{\alpha}})$ must be the spatial coordinate of an orbital of type $\beta$. This implies
\bal \label{pgscond2}
U(g)_{\beta \alpha} \neq 0 \imp \boldsymbol{\Delta}_{\beta \alpha} \in \text{BL}.
\end{align}
For example, consider a basis of $(p_x,p_y)$ orbitals in a 2D monoatomic square lattice. We choose that the spatial origin coincides with the position of one atom, thus the spatial embedding of the orbitals are $r_{p_x}=r_{p_y}=0$. We employ the shorthand that $(x,y)$ represents a vector $x \hat{x} + y \hat{y}$. A four-fold rotation $(g=C_4)$ transforms vectors as $(x,y) \rightarrow  (-y,x)$, thus it is represented in $\R^2$ by $D(C_4) = -i \sigma_2$, with $\sigma_2$ a Pauli matrix.   Since $(p_x,p_y)$ orbitals also transform in the vector representation, we find that $C_4$ is represented in the orbital basis by $U(C_4) = -i \sigma_2$. A square lattice is symmetric under $C_4$, thus for any BL vector $\boldsymbol{R}$, $D(C_4)\boldsymbol{R}$ is also a BL vector. In a monatomic BL where the spatial origin coincides with the atom, $\boldsymbol{\Delta}_{\ab}=0$ trivially.\\ 

Applying (\ref{blochbasis}), (\ref{pgscond1}), (\ref{pgscond2}) and the orthogonality of $D(g)$, the Bloch basis vectors transform as
\bal
\hat{g}\,\dg{c}_{\boldsymbol{k},\alpha}\,\hat{g}^{\mo}  = \dg{c}_{D(g)\boldsymbol{k},\beta}\,U(g)_{\beta \alpha}.
\end{align}
If the Hamiltonian is symmetric under $g$, $[H,\hat{g}]=0$ implies
\bal
U(g)\,H(\boldsymbol{k})\,U(g)^{\mo} = H\big(\,D(g)\boldsymbol{k}\,\big).
\end{align}

\subsection{Little group of the wavevector} 

Suppose special momenta ($\boldsymbol{\bar{k}}$) exist that satisfies
\bal \label{invmom}
D(g)\boldsymbol{\bar{k}} = \boldsymbol{\bar{k}}+ \boldsymbol{G}_g(\boldsymbol{\bar{k}})
\end{align}
for some reciprocal lattice (RL) vector $\boldsymbol{G}$ that depends on the momentum and the symmetry element in question. We say that $\boldsymbol{\bar{k}}$ as invariant under $g$. We deduce from (\ref{aperiodic}) that
\bal
[\,H(\boldsymbol{\bar{k}}),\,V\big(\,\boldsymbol{G}_g(\boldsymbol{\bar{k}})\,\big)\,U(g)\,]=0.
\end{align}
Equivalently, the eigenstates of $H(\boldsymbol{\bar{k}})$ may be chosen to have quantum numbers under the unitary operation $V\big(\,\boldsymbol{G}_g(\boldsymbol{\bar{k}})\,\big)\,U(g)$. The collection of all symmetry elements $\{g_1,g_2, \ldots, g_l\}$ which leave $\boldsymbol{\bar{k}}$ invariant forms the little group of the wavevector; the little group is generically a subgroup of the group of the Hamiltonian.\cite{tinkhambook} Henceforth, we shall be discussing a single momentum $\boldsymbol{\bar{k}}$, and we suppress writing $\boldsymbol{\bar{k}}$ in the arguments of $\boldsymbol{G}_g$. The set of operations $\{V(\boldsymbol{G}_{g_1})U(g_1), \ldots, V(\boldsymbol{G}_{g_l})U(g_{l(\boldsymbol{\bar{k}})}) \}$ form a representation of the little group at $\boldsymbol{\bar{k}}$. The little group at $\boldsymbol{\bar{k}}=0$ is the group of the Hamiltonian, for which $\{U(g_1), \ldots, U(g_{l(0)}) \}$ form a representation; in this case $V(\boldsymbol{G}_{g})=I$ trivially for all $g$.

Let's introduce the shorthand: $D_a = D(g_a)$ and $U(g_a) = U_a$. A useful identity is
\bal \label{useful}
U_a\,V(\boldsymbol{G})\,U_a^{\mo} = V(D_a\,\boldsymbol{G})
\end{align}
for any reciprocal lattice vector $\boldsymbol{G}$. Proof: applying (\ref{pgscond1}) and (\ref{pgscond2}), we deduce that
\bal
[U_a]_{\ab} \neq 0 \imp D_a^{\mo}\,\boldsymbol{\Delta}_{\ab} \in \text{BL}.
\end{align}
Applying this in conjunction with the orthogonality of $D_a$, we find
\bal
[U_a\,V(\boldsymbol{G})]_{\ab} =  [U_a]_{\ab}\,e^{i\boldsymbol{G}\cdot \boldsymbol{r_{\beta}}}  = e^{i(D_a\boldsymbol{G})\cdot \boldsymbol{r_{\alpha}}}\,[U_a]_{\ab}\,e^{i\boldsymbol{G} \cdot D_a^{\mo}\,\boldsymbol{\Delta}_{\ab}} = [V(D_a\boldsymbol{G})\,U_a\,]_{\ab}.
\end{align}

For a less trivial example of a litte group, we consider the $C_4$-invariant point $\boldsymbol{\bar{k}}=(\pi,\pi,0)$ for spinless $C_{4v}$ systems. Each element $g_a$ in this group is represented by $X(g_a) = V(\boldsymbol{G}_{g_a})\,U(g_a)$:
\bal
&X(e) = I, &&X(C_4) = V(-2\pi \hat{x})\,U(C_4), && X(C_4^{\mo})= V(-2\pi \hat{y})\,U(C_4^{\mo}), \lin
&X(C_2)= V\big(-2\pi (\hat{x}+\hat{y})\,\big)\,U(C_2), &&X(M_x)=V(-2\pi \hat{x})\,U(M_x), &&X(M_y)=V(-2\pi \hat{y})\,U(M_y), \lin
&X(M_1) = U(M_1), &&X(M_2)= V\big(-2\pi (\hat{x}+\hat{y})\,\big)\,U(M_2), &&
\end{align}
where $e$ is the identity element, $C_n$ is an $n$-fold rotation, and $M_i$ are reflections which transform real-space coordinates as $M_x: (x,y) \rightarrow (-x,y)$, $M_y: (x,y) \rightarrow (x,-y)$, $M_1: (x,y) \rightarrow (y,x)$, $M_2: (x,y) \rightarrow (-y,-x)$. Applying (\ref{useful}), one derives that these matrices satisfy the requisite algebraic relations, \emph{e.g.}, $X(C_4)^4 = X(e)$, $X(M_i)^2=X(e)$ and $X(M_i)\,X(C_4)\,X(M_i)^{\mo} = X(C_4^{-1})$. The last relation is merely the matrix representation of a simple statement: the handedness of a rotation inverts under a reflection, if the rotation axis is parallel to the reflection plane.

\subsection{Generalized little groups} \label{app:groupMP}

Let us consider lower-dimensional submanifolds which are embedded in the 3D BZ. The set of all elements which leave this submanifold invariant is defined as the little group of the submanifold. The little group of the wavevector corresponds to a 0D submanifold, but we will also be interested in 1D and 2D submanifolds. For example, let us define a mirror plane (MP) as a plane in the 3D BZ which is mapped to itself under a certain reflection, up to a translation by a RL vector. For example, the plane $k_x=\pi$ is mapped to itself under the reflection $M_x: (x,y) \rightarrow (-x,y)$, up to a translation of $\boldsymbol{G}=2\pi \hat{x}$. We define the group of the MP as the collection of all symmetry elements $\{g_1,g_2, \ldots, g_l\}$ which leave MP invariant; the group of the MP is generically a subgroup of the group of the Hamiltonian, which we take to be $\cfv$ for illustration. In the spinless representation, the group of the MP $k_x=\pi$ consists of the elements $\{e,M_x,C_2,M_y\}$. We recall the definitions of $U(g)$, $D(g)$ and $V(\boldsymbol{k})$ in App. \ref{app:rep}. Suppose $\boldsymbol{k} \in MP$, $\boldsymbol{G}_g(\text{MP})$ is defined as the RL vector that separates $D(g)\boldsymbol{k}$ from the MP:
\bal
\forall \;\;\boldsymbol{k},\boldsymbol{k'} \in MP, \;\;\;D(g)\boldsymbol{k} = \boldsymbol{k'} + \boldsymbol{G}_g(\text{MP}).
\end{align}
Each element in the group of the MP is represented by $X(g) = V(\boldsymbol{G}_g(\text{MP}))\,U(g)$. For example, the group of the MP $k_x=\pi$ is represented by
\bal
&X(e) = I, &&X(C_2) = V(-2\pi \hat{x})\,U(C_2), && X(M_x)=V(-2\pi \hat{x})\,U(M_x), &&X(M_y)=U(M_y).
\end{align}
One may verify through the identity (\ref{useful}) that these matrices satisfy the requisite algebraic relations, \emph{e.g.}, $X(C_2)^2=X(e)$, and  $X(M_x)\,X(C_2)\,X(M_x^{\mo}) = X(C_2^{\mo}) = X(C_2)$.

\section{Mirror Chern numbers in systems with $C_{nv}$ symmetry} \label{app:vanishmc}

\subsection{Definition of mirror Chern numbers, in systems with or without spin-orbit coupling}

If an energy gap exists that distinguishes between occupied and unoccupied bands, we may define the Berry vector potential as
\bal \label{defvecpot}
\cala(\boldsymbol{k}) = -i\,\sum_{n\in occ.} \,\bra{u_{n,\boldsymbol{k}}}\,\nabla\,\ket{u_{n,\boldsymbol{k}}},
\end{align}
and the Abelian Berry field as
\bal 
{\cal \tilde{F}}(\boldsymbol{k}) = \nabla \times \cala(\boldsymbol{k}).
\end{align}  
In (\ref{defvecpot}), we sum over all occupied bands. In $\cnv$ systems, there exist planes in the 3D BZ which are invariant under a certain reflection, up to translations by a reciprocal lattice vector. In each mirror plane (MP$_i$), there exists an operator $X(M_i)$ which represents the reflection $M_i$; $X(M_i)$ represents an element in the group of the mirror plane, as defined in App. \ref{app:groupMP}. In representations with spin, $X(M_i)^2=-I$ and the eigenvalues of reflection are $\pm i$; we define the mirror-even (-odd) bands as having mirror eigenvalues $+i$ ($-i$). In representations without spin, $X(M_i)^2=+I$ and we define the mirror-even (-odd) bands as having mirror eigenvalues $+1$ ($-1$). Mirror-even bands are denoted by the superscript $(e)$, and we may define the mirror-even Berry field ${\cal \tilde{F} }(\boldsymbol{k})_e$ as 
\bal
{\cal \tilde{F} }(\boldsymbol{k})_e = -i\sum_{n\in occ,even} \nabla \times  \,\bra{u^e_{n,\boldsymbol{k}}}\,\nabla\,\ket{u^e_{n,\boldsymbol{k}}},
\end{align}  
where we only sum over occupied bands which transform in the even representation of reflection. We similarly define the mirror-odd Berry field ${\cal \tilde{F} }(\boldsymbol{k})_o$. In each MP$_i$ we denote an infinitesimal, directed area element by $d\Omega$. The even and odd mirror Chern numbers are defined as
\bal \label{mircherndef}
\calc_e = \frac{1}{2\pi}\int_{\text{MP}} d\Omega \cdot {\cal \tilde{F} }(\boldsymbol{k})_e,\;\; \calc_o = \frac{1}{2\pi}\int_{\text{MP}} d\Omega \cdot {\cal \tilde{F} }(\boldsymbol{k})_o.
\end{align}

\subsection{The mirror Chern numbers vanish for spinless systems, with either $\calc_{2v}, \cfv$ or $\csv$ symmetry} \label{app:vanishsu2}

By spinless systems, we refer either to electronic systems with spin $SU(2)$ symmetry, or to intrinically spinless systems such as photonic crystals and certain cold atoms.  For $\cnv$ systems with $n=2,4$ or $6$, there exists a two-fold rotational symmetry about the $\hat{z}$ axis. As shown in App. \ref{app:rep}, this symmetry manifests as
\bal
U(C_2)\,H(\boldsymbol{k})\,U(C_2)^{\mo} = H\big( \,D(C_2)\cdot \boldsymbol{k} \,\big),
\end{align}
where $U(C_2)$ represents a two-fold rotation in the orbital basis, and 
\bal
D(C_2) = \begin{pmatrix} -1 & 0 & 0 \\
                         0 &  -1 & 0 \\
                         0&  0& +1 \end{pmatrix}
                         \end{align}
                         represents a two-fold rotation in $\R^3$. Since the rotation axis lies within the mirror plane MP$_i$, the element $C_2$ leaves MP$_i$ invariant, thus $C_2$ belongs to the little group of MP$_i$. As shown in App. \ref{app:groupMP}, each element $g$ in this subgroup  is represented by  an operator $X(g)$, which satisfy requisite algebraic relations that follows from the group structure, \emph{e.g.}, $X(M_i)\,X(C_2)\,X(M_i^{\mo}) = X(C_2^{\mo})$. Since these representations are assumed to be spinless, $X(C_2)^2=I$, and $X(C_2)$ commutes with $X(M_i)$. This implies that a mirror-even state at $\boldsymbol{k}$ is mapped to a mirror-even state at $D(C_2)\boldsymbol{k}$, by two-fold symmetry. It follows that the mirror Berry fields are related by
\bal \label{fieldrel1}
{\cal \tilde{F} }(\boldsymbol{k})_e = D(C_2) \cdot {\cal \tilde{F} } \big( \,D(C_2)\cdot \boldsymbol{k} \,\big)_e, \;\;\text{and}\; {\cal \tilde{F} }(\boldsymbol{k})_o = D(C_2) \cdot {\cal \tilde{F} } \big( \,D(C_2)\cdot \boldsymbol{k} \,\big)_o.
\end{align}
There exists Euclidean basis vectors $(\hat{e}_{\parallel,1},\hat{e}_{\parallel,2},\hat{e}_{\perp})$ which transform under the reflection into $(\hat{e}_{\parallel,1},\hat{e}_{\parallel,2},-\hat{e}_{\perp})$; the subscript $\parallel$ ($\perp$) denotes a vector that is parallel (perpendicular) to the mirror plane. Since the two-fold rotational axis lies within MP$_i$, $D(C_2) \cdot \hat{e}_{\perp} = -\hat{e}_{\perp}$, and 
\bal
{\cal \tilde{F} }(\boldsymbol{k})_e \cdot \hat{e}_{\perp} = - {\cal \tilde{F} } \big( \,D(C_2)\cdot \boldsymbol{k} \,\big)_e \cdot \hat{e}_{\perp}.
\end{align}
Since the directed area element $d\Omega$ is parallel to $\hat{e}_{\perp}$, the contributions to the integral (\ref{mircherndef}) at $\boldsymbol{k}$ and $D(C_2) \boldsymbol{k}$ are equal in magnitude but opposite in sign. This implies $\calc_e=0$, and a similar argument can be made for $\calc_o=0$. It should be noted that in spin-orbit-coupled systems, $X(M_i)$ \emph{anti}commutes with $X(C_2)$, thus $\calc_e=-\calc_o$ instead.

\subsection{Mirror Chern numbers in spinless $\ctv$ systems}

In $\ctv$ systems without spin-orbit coupling, two-fold rotational symmetry is absent, thus the mirror Chern numbers need not vanish for the reason stated in Sec. \ref{app:vanishsu2}. These invariants might vanish for other reasons: if the $\ctv$ system belongs to a larger symmetry group, for which $\ctv$ is a subgroup, then $\calc_e=\calc_o=0$ if there exists either of these additional symmetries: (a) a reflection plane that is orthogonal to the principal $C_3$ axis, or (b) a two-fold axis that lies perpendicular to the $C_3$ axis, and parallel to the mirror-plane. The proof is very similar to that in App. \ref{app:proofvanish1} and \ref{app:proofvanish2}.

\section{$\boldsymbol{k} \cdot \boldsymbol{p}$ analysis of surface bands in spinless $\cnv$ systems} \label{app:kdotp}

Consider a $C_n$-invariant point which is contained in a mirror line in the 001 surface BZ -- the little group of the wavevector is $\cnv$.  We choose a coordinate system such that: (i) the origin lies at the $C_n$-invariant point, (ii) $\hat{z}$ lies along the principal $C_n$ axis, and (ii) $\hat{x}$ is parallel to the mirror line, \emph{i.e.}, the reflection $M_y$ transforms $(x,y) \rightarrow (x,-y)$. We consider surface bands that transform in the doublet irrep of $\cnv$. Each doublet irrep comprises two states with distinct, complex-conjugate eigenvalues under an $n$-fold rotation; these two states are degenerate because $M_y\,C_n\,M_y^{\mo} = C_n^{\mo}$. It should be noted that there are no complex eigenvalues under a two-fold rotation, thus our discussion applies to $n>2$. Applying $C_n^n=e$ (the identity), we deduce these eigenvalues: $(e^{i2\pi/3},e^{-i2\pi/3})$ for $\ctv$, and $(i,-i)$ for $\cfv$. For $\csv$ there are two pairs:  $(e^{i2\pi/3},e^{-i2\pi/3})$ and $(e^{i\pi/3},e^{-i\pi/3})$. In summary, there is one doublet irrep in each of the groups $\ctv$ and $\cfv$, and two doublet irreps in the group $\csv$. 

\subsection{Doublet irreducible representation of type 1} \label{type1}

For the groups $\ctv, \cfv$ and $\csv$, there exists a doublet irrep which transforms as vectors $(x,y)$. We choose a two-dimensional basis in which $\ket{1}$ transforms as $x+iy$ and $\ket{2}$ as $x-iy$. In this basis, the representations of our symmetry elements are
\bal
U(M_y) = \sigma_1, \;\;\text{and}\;\; U(C_n) = e^{i 2\pi \sigma_3/n}.
\end{align}
The effective two-band Hamiltonian may be expressed as
\bal
H(\boldsymbol{k}) = d(\boldsymbol{k})\,I + f(\boldsymbol{k})\,\sigma_+ + f(\boldsymbol{k})^*\,\sigma_- + g(\boldsymbol{k})\,\sigma_3,
\end{align}
where $d(\boldsymbol{k})$ and $g(\boldsymbol{k})$ are real functions, $f(\boldsymbol{k})$ is generally complex, and $\sigma_{\pm} = \sigma_1 \pm i \sigma_2$. This Hamiltonian satisfies the symmetry relations
\bal
& U(M_y)\,H(k_+,k_-)\,U(M_y^{\mo}) = H\big(\,k_-,k_+\,\big), \\
& U(C_n)\,H(k_+,k_-)\,U(C_n^{\mo}) = H\big(\,k_+\,\omega, \,k_-\,\omega^*\big),
\end{align}
where $k_{\pm} = k_x \pm i k_y$ and $\omega= \text{exp}[i 2\pi/n]$. We expand 
\bal
& d(\boldsymbol{k}) = \sum_{i\geq 0,j\geq 0} d_{ij} \,k_+^i\,k_-^j;\;\;\; f(\boldsymbol{k}) = \sum_{i\geq 0,j\geq 0} f_{ij} \,k_+^i\,k_-^j ;\;\;\; g(\boldsymbol{k}) = \sum_{i\geq 0,j\geq 0} g_{ij} \,k_+^i\,k_-^j.
\end{align}
$C_n$ symmetry imposes
\bal
&d_{ij} = 0 \;\;\text{if}\;\; \frac{i-j}{n} \notin \Z ;\;\;\;f_{ij} = 0 \;\;\text{if}\;\; \frac{i-j-2}{n} \notin \Z ;\;\;\;g_{ij} = 0 \;\;\text{if}\;\; \frac{i-j}{n} \notin \Z.
\end{align}
$M_y$ symmetry imposes
\bal \label{same}
& d_{ij} = d_{ji} ;\;\;\; f_{ij} \in \R ;\;\;\; g_{ij} = -g_{ji}.
\end{align}
If the group of the wavevector is $\ctv$, an expansion of the effective Hamiltonian to second order in $k$ gives
\bal
& d(\boldsymbol{k}) = m+ d \,k_+\,k_-;\;\;\; f(\boldsymbol{k}) = a\,k_- + b\,k_+^2;\;\;\; g(\boldsymbol{k}) = 0.
\end{align}
For $\cfv$, 
\bal
& d(\boldsymbol{k}) = m+ d \,k_+\,k_- ;\;\;\; f(\boldsymbol{k}) = a\,k_-^2 + b\,k_+^2;\;\;\; g(\boldsymbol{k}) = 0.
\end{align}
For $\csv$, 
\bal
& d(\boldsymbol{k}) = m+ d \,k_+\,k_- ;\;\;\; f(\boldsymbol{k}) = b\,k_+^2;\;\;\; g(\boldsymbol{k}) = 0.
\end{align}
All coefficients $\{m,d,a,b\}$ are real.

\subsection{Doublet irreducible representation of type 2}

For the group $\csv$, there exists a second doublet irrep which transforms as $(x^2-y^2,2xy)$. We choose the same coordinates as in Sec. \ref{type1}, and a two-dimensional basis in which $\ket{1}$ transforms as $(x^2-y^2)+i2xy$ and $\ket{2}$ as $(x^2-y^2)-i2xy$. In this basis, the representations of our symmetry elements are
\bal
U(M_y) = \sigma_1, \;\;\text{and}\;\; U(C_6) = e^{i 2\pi \sigma_3/3}.
\end{align}
The two-band effective Hamiltonian satisfies the symmetry relations
\bal
& U(M_y)\,H(k_+,k_-)\,U(M_y^{\mo}) = H\big(\,k_-,k_+\,\big), \\
& U(C_6)\,H(k_+,k_-)\,U(C_6^{\mo}) = H\big(\,k_+\,\omega, \,k_-\,\omega^*\big),
\end{align}
where $k_{\pm} = k_x \pm i k_y$ and $\omega= \text{exp}[i 2\pi/6]$. $C_6$ symmetry imposes
\bal
&d_{ij} = 0 \;\;\text{if}\;\; \frac{i-j}{6} \notin \Z ;\;\;\;f_{ij} = 0 \;\;\text{if}\;\; \frac{i-j-4}{6} \notin \Z ;\;\;\;g_{ij} = 0 \;\;\text{if}\;\; \frac{i-j}{6} \notin \Z.
\end{align}
$M_y$ symmetry imposes the same constraints as in (\ref{same}). An expansion of the effective Hamiltonian to second order in $k$ gives
\bal
& d(\boldsymbol{k}) = m+ d \,k_+\,k_-;\;\;\; f(\boldsymbol{k}) = a\,k_-^2 ;\;\;\; g(\boldsymbol{k}) = 0.
\end{align}
All coefficients $\{m,d,a\}$ are real.

\section{The halved mirror chirality} \label{app:halved}

\subsection{Integer quantization of the halved mirror chirality} \label{app:proofinteger}

Let us prove that $\chi \in \Z$, for systems with or without spin-orbit coupling. From the definition (\ref{mirrorchirality}), we separate $\chi$ into two parts: $2\pi\,\chi = B_e - B_o$, where
\bal \label{line}
B_{\eta} = \int_{HMP} dt\,dk_z\,\calf_{\eta}(t,k_z),
\end{align}
and $\eta \in \{e,o\}$ distinguishes the mirror-even and mirror-odd subspaces, as defined in App. \ref{app:vanishmc}. The Berry curvature in a mirror subspace is defined by
\bal \label{definemirrorcurv1}
\calf_{\eta}(t,k_z) = \partial_t\,\cala_{z}^{\eta}(t,k_z) -  \partial_z \,\cala_{t}^{\eta}(t,k_z)
\end{align}
where $\partial_t = \partial/\partial t$, $\partial_z = \partial/\partial k_z$, and 
\bal \label{definemirrorcurv2}
\cala_{\mu}^{\eta}(t,k_z) =-i\,\sum_{n\in occ,\eta} \,\bra{u_{n,(t,k_z)}^{\eta}}\,\partial_{\mu}\,\ket{u_{n,(t,k_z)}^{\eta}}.
\end{align}
Here we only sum over occupied bands in the representation of reflection denoted by $\eta$. If $\hat{e}_{\perp}$ is the unit vector orthogonal to HMP, then $\calf_{\eta} = {\cal \tilde{F}}_{\eta} \cdot \hat{e}_{\perp}$ is a scalar, in comparison with the vector ${\cal \tilde{F}}_{\eta}$ which is defined in App. \ref{app:vanishmc}.  In terms of $\cala$,
\bal \label{fluxeta}
B_{\eta} = \int_0^{1} dt\,\partial_t \,\int_{-\pi}^{\pi}dk_z\,\cala_{z}^{\eta} -  \int_{-\pi}^{\pi} dk_z \partial_z \,\int_0^1 dt\,\cala_{t}^{\eta}.
\end{align}
Since $B_{\eta}$ is expressed in terms of Berry curvature, it is manifestly gauge-invariant. Given that $\ket{u_{n,(t,k_z)}}$ is an eigenstate of $H(t,k_z)$ in the HMP, it follows from (\ref{aperiodic}) that $V(2\pi \hat{z})^{\mo}\,\ket{u_{n,(t,k_z)}}$ is an eigenstate of $H(t,k_z+2\pi)$, up to a $U(1)$ phase ambiguity. It is convenient to choose this arbitrary phase to vanish as
\bal \label{periodicgauge}
\ket{u_{n,(t,k_z+2\pi)}} = V(2\pi \hat{z})^{\mo}\,\ket{u_{n,(t,k_z)}},
\end{align}
or equivalently $\psi_{n,(t,k_z)}(\boldsymbol{r}) = \psi_{n,(t,k_z+2\pi)}(\boldsymbol{r})$. In this periodic gauge, the second term of (\ref{fluxeta}) vanishes.\cite{kingsmith1993} The remaining expression may be expressed in terms of the Wilson loop, which is a matrix representation of holonomy. Let us consider the parallel transport of a mirror-eigenstate, around a non-contractible loop in the HMP -- at constant $t$ but varying $k_z$. In the orbital basis, such transport is represented by the operator
\bal \label{con1}
\hat{\W}_{\eta}(t) = V(2\pi \hat{z})\,\prod_{k_z}^{\pi \leftarrow -\pi}\,P_{\eta}(t,k_z),
\end{align}
where $P_{\eta}(t,k_z)$ projects into the occupied subspace in the $\eta$-representation of reflection:
\bal \label{con2}
P_{\eta}(t,k_z) = \sum_{n \in occ} \ket{u_{n,(t,k_z)}^{\eta}}\bra{u_{n,(t,k_z)}^{\eta}},
\end{align}
and $(\pi \leftarrow -\pi)$ indicates that the product of projections is path-ordered. In the basis of $n_{occ,\eta}$ occupied bands in the $\eta$-representation, this same parallel transport is represented by an $n_{occ,\eta}\times n_{occ,\eta}$ matrix:
\bal \label{con3}
[\W_{\eta}(t)]_{ij} = \bra{u_{i,(t,-\pi)}^{\eta}}\,\hat{\W}_{\eta}(t)\,\ket{u_{j,(t,-\pi)}^{\eta}},
\end{align}
where the total eigenspectrum of (\ref{con3}) comprise the unimodular eigenvalues of (\ref{con1}). It is shown in Ref. \onlinecite{aa2012}  that (\ref{fluxeta}) may be expressed as 
\bal \label{nice}
B_{\eta} \eq \int_0^{1} dt\,\partial_t \,\int_{-\pi}^{\pi}dk_z\,\cala_{z}^{\eta} = -i\int_0^{1} dt\,\partial_t \,\text{ln}\,\text{det}\,\W_{\eta}(t) \lin
\eq -i\,\text{ln}\,\text{det}\,\W_{\eta}(t=1) +i\,\text{ln}\,\text{det}\,\W_{\eta}(t=0) + 2\pi\,N_{\eta}; \;\;\; N_{\eta} \in \Z.
\end{align}
Here we have chosen the principal branches of the logarithm at endpoints $t=0$ and $t=1$, and $N_{\eta}$ is the number of windings in the interval $t \in (0,1)$, relative to these principal values. We claim that the eigenspectrum of $\W_{\eta}(t=0)$ is degenerate with the eigenspectrum of $\W_{-\eta}(t=0)$, and a similar degeneracy occurs at the other endpoint $t=1$. \\

\noindent Proof: let $\bar{t}$ denote an endpoint $t \in \{0,1\}$. The lines at constant $\bar{t}$ are invariant under (i) a certain $m$-fold rotation ($m=3$ for $\ctv$ and $\csv$, $m=4$ for $\cfv$), and (ii) a reflection $M_i$. The set of elements which leave this line invariant form the group of the line; they are represented by $\{X(M_i),X(C_m),X(C_m^{\mo}), \ldots\}$, as discussed in App. \ref{app:groupMP}. The relevant symmetry relations are
\bal \label{relevant}
\forall \;\;k_z,\;\;\;[\,X(C_m),H(\bar{t},k_z)\,] = [\,X(C_m^{\mo}),H(\bar{t},k_z)\,] = [\,X(M_i),H(\bar{t},k_z)\,] =0.
\end{align}
Analogous to the construction of (\ref{con2}) and (\ref{con3}), we may define, for \emph{all} $n_{occ}$ occupied bands, a $n_{occ} \times n_{occ}$ Wilson-loop matrix:
\bal \label{analogcons}
[\W(\bar{t})]_{ij} = \bra{u_{i,(\bar{t},-\pi)}}\;V(2\pi \hat{z})\,\prod_{k_z}^{\pi \leftarrow -\pi}\,\big(\,P_{\eta}(\bar{t},k_z) + P_{-\eta}(\bar{t},k_z)\,\big) \;\ket{u_{j,(\bar{t},-\pi)}} = \bra{u_{i,(\bar{t},-\pi)}}\;\hat{\W}(t) \;\ket{u_{j,(\bar{t},-\pi)}}.
\end{align}
We also define the complete projection into the occupied bands as $P = P_{\eta} + P_{-\eta}$, which projects onto both mirror-odd and -even bands. Since the bands transform in the doublet irreps, $n_{occ,\eta}+n_{occ,-\eta} = n_{occ}$, as assumed in the main text. Since the mirror subspaces are orthogonal, the eigenspectum of $\W(\bar{t})$ comprise the eigenvalues of $\W_{\eta}(\bar{t})$ and $\W_{-\eta}(\bar{t})$.  Suppose $\ket{\eta}$ belongs to the occupied subspace at momentum $(\bar{t},k_z=-\pi)$, and the state satisfies two conditions: (i) it is an eigenstate of $\W(\bar{t})$: $\hat{\W}(\bar{t})\,\ket{\eta} = e^{i\vartheta}\,\ket{\eta}$, and (ii) it is also an eigenstate of reflection:  $X(M_i)\,\ket{\eta} = \eta \,\ket{\eta}$. It follows from (\ref{relevant}) that $X(C_m)\ket{\eta}$ and $X(C_m^{\mo})\ket{\eta}$ belong in the occupied subspace at $(\bar{t},k_z=-\pi)$. Since $\ket{\eta}$ transforms in the doublet irrep, it is a linear combination of states $\{\ket{\lambda},\ket{\lambda^*}\}$ with complex eigenvalues under $X(C_m)$ (for $m>2$). Note that if $X(C_m)\ket{\lambda} = \lambda \ket{\lambda}$ for complex $\lambda$, then $X(C_m^{\mo})\ket{\lambda} = \lambda^*\ket{\lambda} \neq X(C_m)\ket{\lambda}$. Thus we deduce that $(\,X(C_m)-X(C_m^{\mo})\,)\ket{\eta}$ is not a null vector. Then
\bal
X(M_i)\,X(C_m)\,X(M_i^{\mo}) = X(C_m^{\mo}) \imp  X(M_i)\,(\,X(C_m) - X(C_m^{\mo})\,) \,\ket{\eta} =  -\eta\,(\,X(C_m) - X(C_m^{\mo})\,) \,\ket{\eta} ,
\end{align}
\emph{i.e.}, $\ket{\eta}$ and $(\,X(C_m) - X(C_m^{\mo})\,) \,\ket{\eta}$ have opposite mirror eigenvalues. From (\ref{relevant}), we derive 
\bal
\forall \;\;k_z, \;\;\;X(C_m)\,P(\bar{t},k_z)\,X(C_m^{\mo}) = P(\bar{t},k_z).
\end{align}
Combining this relation with (\ref{useful}) and (\ref{analogcons}), 
\bal
\hat{\W}(\bar{t})\,(\,X(C_m) - X(C_m^{\mo})\,) \,\ket{\eta} = e^{i\vartheta}\,(\,X(C_m) - X(C_m^{\mo})\,) \,\ket{\eta}.
\end{align}
This proves the double-degeneracy in the Wilson-loop spectrum, or equivalently,
\bal \label{deg}
\text{ln}\,\text{det}\,\W_{\eta}(\bar{t}) = \text{ln}\,\text{det}\,\W_{-\eta}(\bar{t}).
\end{align}
Combining this with (\ref{nice}) and (\ref{mirrorchirality}), we find
\bal
\chi_i = \frac{N_{\eta} - N_{-\eta}}{2\pi} = N_{\eta} - N_{-\eta} \in \Z,
\end{align}
where $\eta=1$ for spinless representations, and $i$ for representations with spin.

\subsection{$\chi=0$ with a reflection plane orthogonal to the principal $C_n$ axis} \label{app:proofvanish1}

All Bloch wavefunctions in HMP$_i$ may be diagonalized by a single operator, which represents the reflection $M_i$. Suppose there exists another reflection symmetry $M_z$: $z \rightarrow -z$, for $\hat{z}$ along the principal $C_n$ axis. Since $M_i$ and $M_z$ are reflections in perpendicular planes, their operations commute. This implies that a mirror-even state at $\boldsymbol{k}$ is mapped to a mirror-even state at $D(M_z)\boldsymbol{k}$, hence the mirror Berry curvatures are related by $\calf_{e}(t,k_z) = -\calf_{e}(t,-k_z).$ Similarly, $\calf_{o}(t,k_z) = -\calf_{o}(t,-k_z).$ In comparison with (\ref{fieldrel1}), there is an extra minus sign because the Berry field ${\cal \tilde{F}}$ is a pseudovector, and $D(M_z)$ is an improper rotation. From (\ref{line}), it follows that $B_{e}=B_{o}=\chi=0$.

\subsection{$\chi=0$ with a two-fold axis that lies perpendicular to the principal $C_n$ axis, and parallel to the half-mirror-plane}  \label{app:proofvanish2}

Suppose there exists a two-fold rotational symmetry $C_2$, with axis perpendicular to the principal $C_n$ axis, and parallel to HMP$_i$. For spinless representations, $M_i C_2 M_i^{\mo} = C_2^{\mo} = C_2$, thus their operations commute. This implies that a mirror-even state at momentum $(t,k_z)$ within HMP$_i$ is mapped to a mirror-even state at $(t,-k_z)$ by the two-fold symmetry, hence the mirror Berry curvatures are related by $\calf_{e}(t,k_z) = -\calf_{e}(t,-k_z).$ Similarly, $\calf_{o}(t,k_z) = -\calf_{o}(t,-k_z).$ From (\ref{line}), it follows that $B_{e}=B_{o}=\chi=0$.

\section{Relations between the halved mirror chirality and the bent Chern numbers} \label{app:relatechibent}

\begin{figure}[h]
\centering
\includegraphics[width=8 cm]{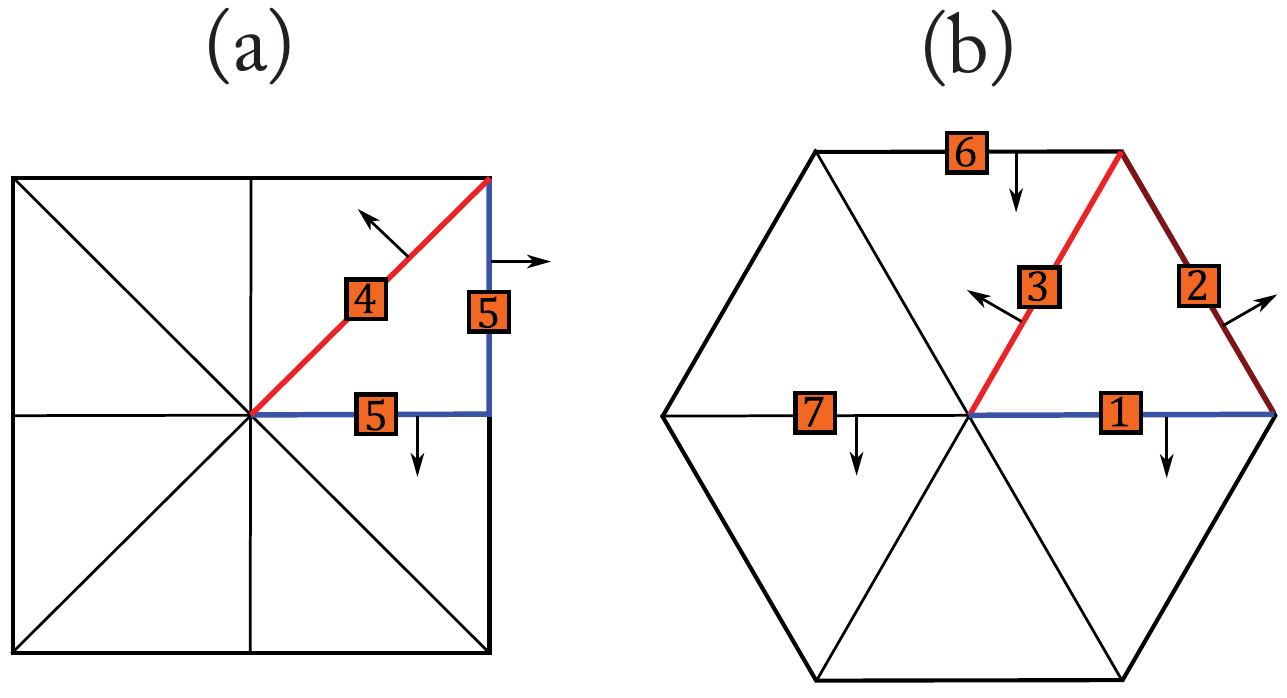}
\caption{Top-down view of 3D BZ's with various symmetries; our line of sight is parallel to the rotational axis. Reflection-invariant planes are indicated by solid lines. Each half-mirror-plane (HMP$_i$) is illustrated by a solid line that connects two distinct $C_m$-invariant lines for $m>2$. HMP$_i$ is labelled by a number $i$ over the solid line. Arrows that emanate from each HMP indicate the convention in which Berry flux is calculated, \emph{i.e.}, they define in vs out. (a) Tetragonal BZ with $\cfv$ symmetry.  (b)  Hexagonal BZ with  $\ctv^{\sma{(b)}}$ symmetry.}\label{suppfig}
\end{figure}

\subsection{The halved mirror chirality and the bent Chern number in $\cfv$ systems} \label{app:c4v}

The halved chiralities and the bent Chern number are related by
\bal \label{parity45app}
\text{parity}[ \,\pdg{\chi}_{4} + \pdg{\chi}_{5}\,] = \text{parity}[\, \calc_{45}\,].
\end{align}
Proof: for $i \in \{4,5\}$, define the mirror Berry flux through HMP$_i$ as
\bal
B_{\eta}^{(i)} = \int_{\text{HMP}_i} dt_i\,dk_z\,\calf_{\eta}(t_i,k_z),
\end{align}
where $\eta \in \{e,o\}$ distinguishes between mirror-even and mirror-odd subspaces; $\calf_{\eta}$ is defined in (\ref{definemirrorcurv1}) and (\ref{definemirrorcurv2}). We choose the convention that this flux emanates from the inside of the triangular pipe, as illustrated in Fig. \ref{suppfig}(a). The total Berry flux is defined $B^{(i)} =  B_{e}^{(i)} + B_{o}^{(i)}$, and the bent Chern number satisfies $2\pi\,\calc_{45} = B^{(4)} + B^{(5)}$. The halved chiralities are defined by $2\pi \chi_i = B_{e}^{(i)} - B_{o}^{(i)}$, thus
\bal
\calc_{45}  = \chi_4 + \chi_5 + \frac{2(\,B_{o}^{(4)} + B_{o}^{(5)}\,)}{2\pi}.
\end{align}
(\ref{parity45app}) follows from proving
\bal \label{toprove}
\frac{B_{o}^{(4)} + B_{o}^{(5)}}{2\pi}\in \Z,
\end{align}
which we now do. As in (\ref{nice}), we express $B$ in terms of the Wilson loop:
\bal \label{binterms}
B^{(i)}_{o} \eq -i\,\text{ln}\,\text{det}\,\W_{M_i=\eta}(t_i=1) +i\,\text{ln}\,\text{det}\,\W_{M_i=\eta}(t_i=0) + 2\pi\,N^{(i)}; \;\;\; N^{(i)} \in \Z.
\end{align}
The subscript $M_i=\eta$ means that $\W$ represents the parallel transport of a state with eigenvalue $\eta$ under $X(M_i)$; $\eta=-1$ ($-i$) for representations without (with) spin. With our chosen flux conventions, the lines labelled by $t_4=0$ and $t_5=1$ coincide, and they project to $\bar{\Gamma}$ in the 001 surface BZ, as illustrated in Fig. 1; the lines $t_4=1$ and $t_5=0$ coincide, and they project to $\bar{M}$ in the surface BZ. We rewrite (\ref{binterms}) as
\bal
B^{(4)}_{o} \eq -i\,\text{ln}\,\text{det}\,\W_{M_4=\eta}(\bar{M}) +i\,\text{ln}\,\text{det}\,\W_{M_4=\eta}(\bar{\Gamma}) + 2\pi\,N^{(4)}; \;\;\; N^{(4)} \in \Z, \lin
B^{(5)}_{0} \eq -i\,\text{ln}\,\text{det}\,\W_{M_5=\eta}(\bar{\Gamma}) +i\,\text{ln}\,\text{det}\,\W_{M_5=\eta}(\bar{M}) + 2\pi\,N^{(5)}; \;\;\; N^{(5)} \in \Z.
\end{align}
Now we will prove that the eigenspectrum of $\W_{M_4=\eta}(\bar{k})$ is identical to that of $\W_{M_5=\eta}(\bar{k})$, for both $\bar{k}= \bar{M}$ and $\bar{\Gamma}$. We recall that $M_4$ reflects $(x,y) \rightarrow (y,x)$ and $M_5$ reflects $(x,y) \rightarrow (x,-y)$, thus the product is a four-fold rotation: $C_4=M_4M_5$, or equivalently their representations satisfy $X(C_4)=X(M_4)X(M_5)$. Suppose $X(M_5)\ket{\eta_5} = \eta_5\ket{\eta_5}$. For spinless representations, it follows from $X(M_4)^2=X(M_5)^2=I$ and $X(C_4)=X(M_4)X(M_5)$, that $(I+X(C_4))\ket{\eta_5}$ is an eigenstate of $X(M_4)$ with eigenvalue $\eta_5$. Moreover, since $\ket{\eta_5}$ transforms in the doublet irrep, $(I+X(C_4))\ket{\eta_5}$ is not a null vector.  Then we can show that
\bal 
\text{ln}\,\text{det}\,\W_{M_4=\eta}(\bar{k}) = \text{ln}\,\text{det}\,\W_{M_5=\eta}(\bar{k}),
\end{align}
in similar fashion to the steps preceding (\ref{deg}). For representations with spin, we consider instead $(I-X(C_4))\ket{\eta_5}$, and arrive at the same conclusion. Finally, (\ref{toprove}) is proven through 
\bal
B^{(4)}_{o} = -i\,\text{ln}\,\text{det}\,\W_{M_5=\eta}(\bar{M}) +i\,\text{ln}\,\text{det}\,\W_{M_5=\eta}(\bar{\Gamma}) + 2\pi\,N^{(4)} =- B^{(5)}_{o} + 2\pi\,N^{(5)} +  2\pi\,N^{(4)}.
\end{align}

\subsection{The halved mirror chirality and the bent Chern number in $\ctv^{\sma{(b)}}$ systems} \label{app:c3v}

In addition to HMP's defined in the main text, it is convenient to define HMP$_6$ and HMP$_7$ as illustrated in Fig. \ref{suppfig}(b). For $i \in \{1,2,3,6,7\}$, define the mirror Berry flux through HMP$_i$ as
\bal \label{definehalfflux}
B_{\eta}^{(i)} = \int_{\text{HMP}_i} dt_i\,dk_z\,\calf_{\eta}(t_i,k_z),
\end{align}
where $\eta \in \{e,o\}$ distinguishes between mirror-even and mirror-odd subspaces; $\calf_{\eta}$ is defined in (\ref{definemirrorcurv1}) and (\ref{definemirrorcurv2}). Our flux conventions are illustrated in the same figure.  In HMP$_i$, all states are diagonalized by the reflection $M_i$, and the various reflection operators are related by a three-fold rotation in coordinates: $M_3 = C_3^{\mo}M_1C_3$, $M_2 = C_3^{\mo}M_3C_3$, etc.   The mirror Chern numbers are defined as
\bal \label{def1}
2\pi\,\calc_{\eta} = B^{(7)}_{\eta} +  B^{(1)}_{\eta} +  B^{(6)}_{\eta}.
\end{align}
The halved mirror chiralities are defined as
\bal \label{def2}
2\pi\,\pdg{\chi}_i = B^{(i)}_{e} -  B^{(i)}_{0}; \;\;\; i \in \{1,2,3\}.
\end{align}
The bent Chern number is defined as
\bal \label{def3}
2\pi\,\calc_{123} = \sum_{\eta \in \{e,o\}}\big(\,B^{(1)}_{\eta} +  B^{(2)}_{\eta} +  B^{(3)}_{\eta}\,\big).
\end{align}
Due to the three-fold rotational symmetry,
\bal \label{def4}
B^{(6)}_{\eta} = B^{(2)}_{\eta}, \;\;\text{and}\;\;B^{(7)}_{\eta} = B^{(3)}_{\eta}.
\end{align}
Combining this with (\ref{def1}) and (\ref{def3}), we find
\bal \label{rel1}
\calc_{123} = \calc_{e} + \calc_{o}.
\end{align}
Combining (\ref{def4}) with (\ref{def2}) and (\ref{def3}), we find
\bal \label{rel2}
\pdg{\chi}_1 + \pdg{\chi}_2 +\pdg{\chi}_3 = \calc_{e} - \calc_{o}.
\end{align}
Once we specify $(\pdg{\chi}_1, \pdg{\chi}_2, \pdg{\chi}_3, \calc_{123})$, the mirror Chern numbers are determined through the relations (\ref{rel1}) and (\ref{rel2}). As a corollary, 
\bal \label{parity123}
\text{parity}[ \,\pdg{\chi}_1 + \pdg{\chi}_2 +\pdg{\chi}_3\,] = \text{parity}[ \,\calc_{123}\,].
\end{align}

\section{Modelling $\ctv^{\sma{(b)}}$ and $\csv$ systems} \label{app:models}

We model a $\ctv^{\sma{(b)}}$ system on a hexagonal lattice composed of two interpenetrating triangular sublattices. Defining $k_1 = \boldsymbol{k} \cdot \boldsymbol{a}_1$ and $k_2= \boldsymbol{k} \cdot \boldsymbol{a}_2$, with $\boldsymbol{a}_1 =(1,0,0 )$ and $\boldsymbol{a}_2=(-1/2,\sqrt{3}/2,0)$, the Hamiltonian is
\bal \label{secondmodel}
H(\boldsymbol{k}) = &\big[\,5/2-\co (k_1+\phi) - \co (k_2+\phi) -\co(k_1+k_2-\phi) -\co (k_z)\,\big]\,\gam{30}  + \gam{10} \lin
& +\bigg\{\;z\,\big[\,e^{i\pi/3}\,\co k_1 +e^{-i\pi/3}\,\co k_2 - \co(k_1+k_2)\,\big]\,\gam{1+} + h.c. \;\bigg\} + \si (k_z)\,\gam{20}
\end{align}
with $\phi = 2\pi/3$. $\gam{1+} = \sigma_1 \otimes (\tau_1+i\tau_2)$. $\sigma_3 = \pm 1$  label the two sublattices, and $\tau_3=\pm 1$ label the $\{p_x \pm ip_y\}$ orbitals.  $C_3$ symmetry manifests as 
\bal \label{cnsym}
X(C_n)\,H(\boldsymbol{k})\,X(C_n)^{\mo} = H(\,D(C_n)\boldsymbol{k}\,),
\end{align}
for $n=3$; $X(C_3) = \sigma_0 \otimes \text{exp}(\,i \,2\pi \,\tau_3/3\,)$, and $D(C_n)$ represents an n-fold rotation in $\R^3$. The reflection symmetries include $\Gamma_{01} \,H(\boldsymbol{k})\,\gam{01} = H(\,R(M_3)\boldsymbol{k}\,)$, where $D(M_3)$ represents a reflection across the mirror plane intersecting HMP$_3$. $z=0.25$ describes a gapped phase with trivial $\{\pdg{\chi}_i\}$. As we tune $z$ from $0.25$ to $0.3$, a Berry dipole nucleates in HMP$_2$, then splits into two monopoles with opposite charge; this semimetallic phase is described by $\pdg{\chi}_1=0,\pdg{\chi}_2=1, \pdg{\chi}_3=0$ and $\calc_{123}=1$; \emph{cf.} (\ref{parity123}). As $z$ is further increased to $0.5$, pairs of monopoles converge on HMP$_1$ and annihilate. The resultant gapped phase satisfies $\pdg{\chi}_1=1,\pdg{\chi}_2=1, \pdg{\chi}_3=0$. This process is depicted in Fig. \ref{weyltrajectories}(f). In our final example, we set $\phi=0$ in (\ref{secondmodel}) so that the Hamiltonian additionally satisfies the symmetry relation (\ref{cnsym}) for $n=6$ and $U(C_6) = \sigma_0 \otimes \text{exp}(\,i \,\pi \,\tau_3/3\,)$. $z=0.5$ describes a trivial gapped $\csv$ phase, and increasing $z$ to $0.75$ produces a gapped phase with $\pdg{\chi}_1=-1$; the intermediate Weyl trajectories are illustrated in Fig. \ref{weyltrajectories}(g).

\section{Additional constraints on topological invariants due to time-reversal symmetry} \label{app:trs}

In spinless systems, time-reversal symmetry (TRS) constrains $\calc_e=\calc_o=0$ in $\ctv^{\sma{(a)}}$ systems, $\pdg{\chi}_1=-\pdg{\chi}_3$ and $\pdg{\chi}_2=\calc_e=\calc_o=0$ in $\ctv^{\sma{(b)}}$ systems; no analogous constraints exist for $\cfv$ or $\csv$. 

\subsection{Time-reversal symmetry in spinless $\ctv^{\sma{(a)}}$ systems} \label{app:trs1}

Let $T$ denote the spinless time-reversal operation, and $M$ denote a reflection in the mirror plane (MP) colored red in Fig. 2(a). Let $\boldsymbol{k}$ be a momentum within MP. Since $[T,M]=0$ and the reflection eigenvalues are real for spinless representations, a state in the even representation at $\boldsymbol{k}$ is mapped by time-reversal to a state in the even representation at $-\boldsymbol{k}$, which also lies in MP. It follows that the mirror Berry curvatures are related by $\calf_{e}(\boldsymbol{k}) = -\calf_{e}(-\boldsymbol{k})$, and the net contribution to the integral in (\ref{mircherndef}) is zero, thus $\calc_e=0$. Similarly, $\calf_{o}(\boldsymbol{k}) = -\calf_{o}(-\boldsymbol{k})$ implies $\calc_o=0$.

\subsection{Time-reversal symmetry in spinless $\ctv^{\sma{(b)}}$ systems} \label{app:trs2}

The proof is similar to that in App. \ref{app:trs1}. Time-reversal ($T$) relates states within HMP$_2$, and imposes $B_e^{(2)} = B_o^{(2)} = 0$, as defined in (\ref{definehalfflux}). A product of time-reversal and a three-fold rotation relates states in HMP$_1$ to states in HMP$_3$, thus $B_e^{(1)} = -B_e^{(3)}$, and $B_o^{(1)} = -B_o^{(3)}$. The conclusion is that $\pdg{\chi}_1=-\pdg{\chi}_3$ and $\pdg{\chi}_2=\calc_e=\calc_o=\calc_{123}=0$.

\subsection{Time-reversal symmetry in spinless $\cfv$ and $\csv$ systems} \label{app:trs3}

The following discussion applies to any HMP$_i$ in either $\cfv$ or $\csv$ systems. A product of time-reversal and a two-fold rotation relates states within the same HMP$_i$ as: $\calf_{\eta}(t_i,k_z) = \calf_{\eta}(t_i,-k_z)$, for both mirror-even and -odd subspaces; the parametrization $(t_i,k_z)$ is defined in the main text. This relation does not constrain any of the above-mentioned topological invariants.

\section{Generalization to spinless superconductors} \label{app:sc}

The mirror Chern numbers, bent Chern numbers and halved mirror chirality are readily generalized to mean-field Hamiltonians in the Bogoliubov-de Gennes (BdG) formalism. If Hermitian, the BdG Hamiltonian has a particle-hole redundancy:
\bal \label{PHS}
P\,H(\boldsymbol{k})\,P^{\mo} = -H(-\boldsymbol{k})
\end{align}
for an antiunitary operator $P$. This relation imposes certain constraints on our topological invariants. If there exists a two-fold rotational symmetry about the principal rotation axis of $\cnv$, then all the described invariants vanish. This situation describes $\cfv$ and $\csv$, and is proven in App. \ref{app:sc1} below. In $\ctv$ systems, the only consequence of (\ref{PHS}) is that $\pdg{\chi}_1 = \pdg{\chi}_3$ for $\ctvb$, as shown in App. \ref{app:sc2}.

\subsection{Vanishing invariants of $\cfv$ and $\csv$} \label{app:sc1}

If the BdG energy spectrum is gapped at momentum $\boldsymbol{k}$, we may define the Berry vector potential as
\bal \label{defvecpotSC}
\cala(\boldsymbol{k}) = -i\,\sum_{E_n<0} \,\bra{u_{n,\boldsymbol{k}}}\,\nabla\,\ket{u_{n,\boldsymbol{k}}},
\end{align}
for $\ket{u_{n,\boldsymbol{k}}}$ an eigenstate of the BdG Hamiltonian $H(\boldsymbol{k})$. Here we sum over all bands with negative energies. The negative-energy Berry field is defined as
\bal 
{\cal \tilde{F}}(\boldsymbol{k}) = \nabla \times \cala(\boldsymbol{k}).
\end{align}  
Analogously, it is convenient to define a positive-energy Berry field:
\bal
{\cal \tilde{G}}(\boldsymbol{k}) = -i\,\sum_{E_n>0} \nabla \times \bra{u_{n,\boldsymbol{k}}}\,\nabla\,\ket{u_{n,\boldsymbol{k}}}.
\end{align}
The particle-hole transformation of (\ref{PHS}) relates a positive-energy state at $\boldsymbol{k}$ to a negative-energy state at $-\boldsymbol{k}$; the negative-energy and positive-energy Berry fields are thus related by 
\bal \label{phsconst}
{\cal \tilde{F}}(\boldsymbol{k}) = - {\cal \tilde{G}}(-\boldsymbol{k}).
\end{align}
A useful relation is that the Berry field of all bands is zero, \emph{i.e.}, 
\bal \label{curvealll}
{\cal \tilde{F}}(\boldsymbol{k}) + {\cal \tilde{G}}(\boldsymbol{k}) =0,
\end{align} 
if the superconductor is gapped at $\boldsymbol{k}$. The proof consists of considering an infinitesimal Wilson loop $\W[l]$ around an area element $d\Omega$, centered at momentum $\boldsymbol{\tilde{k}}$. From Stoke's theorem, 
\bal
\text{exp}\big[\;{i(\;{\cal \tilde{F}}(\boldsymbol{\tilde{k}}) + {\cal \tilde{G}}(\boldsymbol{\tilde{k}})\;) \cdot d\Omega}\;\big] = \text{det} \;\W[l], 
\end{align}
and the discretized Wilson loop has the form
\bal
\W[l]_{ij} = \bra{u_{\boldsymbol{\tilde{k}},i}}\,\prod_{\boldsymbol{q} \in l} P_{all}(\boldsymbol{q})\,\ket{u_{\boldsymbol{\tilde{k}},j}},
\end{align}
where the product of projections are path-ordered around the perimeter $l$ of $d\Omega$. Since $P_{all}$ is the projection onto all bands, by the completeness property it is just the identity. Thus,
\bal
\big(\;{\cal \tilde{F}}(\boldsymbol{\tilde{k}}) + {\cal \tilde{G}}(\boldsymbol{\tilde{k}})\,\big) \cdot d\Omega  = 2\pi \,u;\;\;\;\; u \in \Z.
\end{align}
${\cal \tilde{F}}$ and ${\cal \tilde{G}}$ are bounded if there is no singularity due to a band-touching at $\boldsymbol{\tilde{k}}$. Since $d\Omega$ is infinitesimal, and ${\cal \tilde{F}}+{\cal \tilde{G}}$ bounded, $u=0$. Since this proof works for any orientation of the area element, and for any $\boldsymbol{\tilde{k}}$ where the superconductor is gapped, we have proven (\ref{curvealll}).\\

Combining (\ref{phsconst}) with (\ref{curvealll}),
\bal \label{PHSCONS}
{\cal \tilde{F}}(\boldsymbol{k}) ={\cal \tilde{F}}(-\boldsymbol{k}).
\end{align}
Now consider a momentum $\boldsymbol{k}$ in a mirror plane (MP) of a $\cfv$ or $\csv$ system. All Bloch wavefunctions in MP may be diagonalized by a single operator, which represents the reflection $M_i$. Since the particle-hole transformation commutes with $M_i$, (\ref{PHSCONS}) implies $\calf_e(\boldsymbol{k}) =\calf_e(-\boldsymbol{k})$, where $\calf_e$ is the component of ${\cal \tilde{F}}_e$ perpendicular to MP. Suppose there exists a two-fold rotational symmetry $C_2$, with axis parallel to the principal $C_n$ axis. For spinless representations, $M_i C_2 M_i^{\mo} = C_2^{\mo} = C_2$, thus their operations commute. This implies that a mirror-even state at momentum $\boldsymbol{k}$ is related by two-fold symmetry to a mirror-even state at $D(C_2)\boldsymbol{k}$, where $D(C_2)$ is the representation of the two-fold rotation in $\R^3$. This implies $\calf_e(\boldsymbol{k}) =-\calf_e(-D(C_2)\boldsymbol{k})$. By the same argument we deduce that $\calf_e(t_i,k_z) =-\calf_e(t_i,-k_z)$, for a momentum $(t_i,k_z)$ in a half-mirror-plane (HMP$_i$), and similarly $\calf_o(t_i,k_z) =-\calf_o(t_i,-k_z)$ for the odd subspace. This implies $B_e^{(i)} = B_o^{(i)} = 0$, as defined in (\ref{definehalfflux}). Then the halved chirality is zero because $2\pi \chi_i =  B_e^{(i)} -B_o^{(i)}=0$. By similar arguments, we may derive that the mirror Chern numbers and the bent Chern numbers vanish.

\subsection{${\chi}_1 = {\chi}_3$ in spinless $\ctvb$ superconductors} \label{app:sc2}

We refer to Fig. \ref{suppfig}(b). A product of particle-hole transformation and a three-fold rotation relates states in HMP$_1$ to states in HMP$_3$, thus $B_e^{(1)} = B_e^{(3)}$, and $B_o^{(1)} = B_o^{(3)}$. This implies $\pdg{\chi}_1=\pdg{\chi}_3$. There are no constraints on the other invariants of $\ctvb$. 

\end{widetext}

\bibliography{TI-references-2013oct}

\end{document}